\documentclass[pra,twocolumn,preprintnumbers,superscriptaddress]{revtex4-1} 

\usepackage{graphicx}
\usepackage[small]{caption}
\usepackage{subcaption}
\usepackage{ragged2e}
\DeclareCaptionJustification{justified}{\justifying}
\usepackage{latexsym,epsfig,amssymb,amsfonts,amsmath,graphicx,bbm,bm}

\usepackage{color}
\definecolor{dred}{rgb}{.8,0.2,.2}
\definecolor{dyellow}{rgb}{.7,.7,.0}
\definecolor{ddred}{rgb}{.4,.0,.0}
\definecolor{dblue}{rgb}{.2,.2,.8}
 
\newcommand{\ii}{\mathrm{i}}
\newcommand{\0}{\mathbf{0}}

\newcommand{\RR}{\mathbf{R}}
\newcommand{\RRR}{\mathcal{R}}
\newcommand{\AAA}{\mathbf{A}}

\newcommand{\dd}{\; \mathrm{d}}

\newcommand{\ket}[1]{ |  #1 \rangle}

\newcommand{\bra}[1]{ \langle #1  |}

\newcommand{\brakets}[2]{\langle\, #1\,|\,#2\,\rangle}

\newcommand{\brackets}[3]{\langle #1 | #2 | #3 \rangle}

\newcommand{\od}[2]{\frac{\mathrm{d} #1}{\mathrm{d} #2}}

\newcommand{\eqr}[1]{Eq.\ (\ref{#1})}
\newcommand{\tbr}[1]{Table \ref{#1}}
\newcommand{\fir}[1]{Fig.\ \ref{#1}}
\newcommand{\secr}[1]{Sec.\ \ref{#1}}

\newcommand{\an}[1]{\hat{#1}}
\newcommand{\cre}[1]{\hat{#1}^\dag}

\newcommand{\ee}{\textrm{e}}

\newcommand{\kk}{\mathbf{k}}

\newcommand{\rr}{\mathbf{r}}

\newcommand{\vv}{\mathbf{v}}
\newcommand{\zz}{\mathbf{z}}

\newcommand{\SSS}{\mathbf{S}}

\newcommand{\GG}{\mathbf{G}}

\newcommand{\HH}{\mathbf{H}}

\newcommand{\Pii}{\mathbf{\Pi}}
\newcommand{\nablaa}{\boldsymbol{\nabla}}

\begin{document}

\title{Hubbard model for atomic impurities bound by the vortex lattice of a rotating BEC}

\author{T. H. Johnson}
\affiliation{Centre for Quantum Technologies, National University of Singapore, 117543 Singapore}
\affiliation{Clarendon Laboratory, University of Oxford, Parks Road, Oxford OX1 3PU, United Kingdom}
\affiliation{Keble College, University of Oxford, Parks Road, Oxford OX1 3PG, United Kingdom}
\author{Y. Yuan}
\affiliation{Beijing Computational Science Research Center, Beijing 100094, China}
\affiliation{Department of Mathematics, National University of Singapore, 119076 Singapore}
\affiliation{College of Mathematics and Computer Science, Hunan Normal University, Changsha, Hunan Province, China}
\author{W. Bao}
\email{bao@math.nus.edu.sg} 
\affiliation{Department of Mathematics, National University of Singapore, 119076 Singapore}
\author{S. R. Clark}
\email{s.r.clark@bath.ac.uk} 
\affiliation{Department of Physics, University of Bath, Claverton Down, Bath BA2 7AY, United Kingdom}
\affiliation{Keble College, University of Oxford, Parks Road, Oxford OX1 3PG, United Kingdom}
\author{C. Foot}
\affiliation{Clarendon Laboratory, University of Oxford, Parks Road, Oxford OX1 3PU, United Kingdom}
\author{D. Jaksch}
\affiliation{Clarendon Laboratory, University of Oxford, Parks Road, Oxford OX1 3PU, United Kingdom}
\affiliation{Centre for Quantum Technologies, National University of Singapore, 117543 Singapore}
\affiliation{Keble College, University of Oxford, Parks Road, Oxford OX1 3PG, United Kingdom}


\begin{abstract}
We investigate cold bosonic impurity atoms trapped in a vortex lattice formed by condensed bosons of another species. We describe the dynamics of the impurities by a bosonic Hubbard model containing occupation-dependent parameters to capture the effects of strong impurity-impurity interactions. These include both a repulsive direct interaction and an attractive effective interaction mediated by the BEC. The occupation dependence of these two competing interactions drastically affects the Hubbard model phase diagram, including causing the disappearance of some Mott lobes.
\end{abstract}

\maketitle 

Wherever vortices have been detected, there has been interest in particles bound inside them. For example, particles bound in the vortices of rotating superfluid helium~\cite{Cade1965,Douglass1966,Pratt1969,Gamota1970} were used to count~\cite{Packard1969} and visualize~\cite{Williams1974,Bewley2006,Gomez2014} vortices, and determine their properties~\cite{Douglass1968,Glaberson1968}. Meanwhile, bound particle-antiparticle pairs in the vortex lattices of clean type-II superconductors~\cite{Sonier2000} have received theoretical~\cite{Caroli1964,Kramer1975,Gygi1991} and experimental~\cite{Hess1989} attention due to their importance for charge transport~\cite{Rainer1996} and relaxation at low temperatures.

In this Letter, we consider a vortex lattice of a rotating atomic Bose-Einstein condensate (BEC)~\cite{Abo-Shaeer2001,Madison2000,Schweikhard2004,Williams2010,Fetter2009} in which a small number of cold bosonic atoms, called impurities, are bound. 
Imbalanced cold atomic mixtures have been used to study, experimentally and theoretically, the effect of the majority species on the transport of the impurities~\cite{Palzer2009,Mathy2012,Catani2012,Johnson2012}, often with an external lattice potential~\cite{Ponomarev2006,Bruderer2010,Johnson2011}, and the formation of polarons~\cite{Bruderer2007,Klein2007,Schirotzek2009}. Reference~\cite{Caracanhas2013} considered the continuum modes of impurities immersed in a BEC vortex lattice, but not the regime in which the bound modes of impurities are important.

For this regime, we develop a Hubbard model description for the impurities. To account for strong repulsive interactions between impurities~\cite{Li2006,Johnson2009,Hazzard2010,Buchler2010,Buchler2010e,Dutta2011,Bissbort2012,Luhmann2012,Lacki2013,Major2014}
and with the bosons comprising the BEC~\cite{Luhmann2008,Lutchyn2009,Mering2011,Jurgensen2012}, we allow the wavefunctions of particles at each site to depend on the number of impurities at that site, leading to an occupation-dependent Hubbard model~\cite{Dutta2014}. Our system contrasts with typical experiments featuring cold atoms in optical lattices. The softness of the lattice and the interaction of impurities with lattice degrees of freedom is intrinsic to the system, similar to solid-state systems described by so-called dynamical Hubbard models~\cite{Hirsch1989,Strack1993,Hirsch1994,Amadon1996,Kaiser2014}, including both local distortions and long-ranged degrees of freedom.

We focus on the on-site interactions between impurities, which govern the strongly-interacting part of the phase diagram. We find that occupation dependence, in conjunction with competition between direct repulsive impurity-impurity interactions and effective attractive interactions mediated by the BEC~\cite{Bruderer2007,Klein2007}, drastically alters the typical structure of the ground state phase diagram for these systems. We give an example in which low-occupation Mott lobes are missing entirely.


\begin{figure*}[t]
		\begin{subfigure}[c]{4.2cm}
                \includegraphics[width=\textwidth]{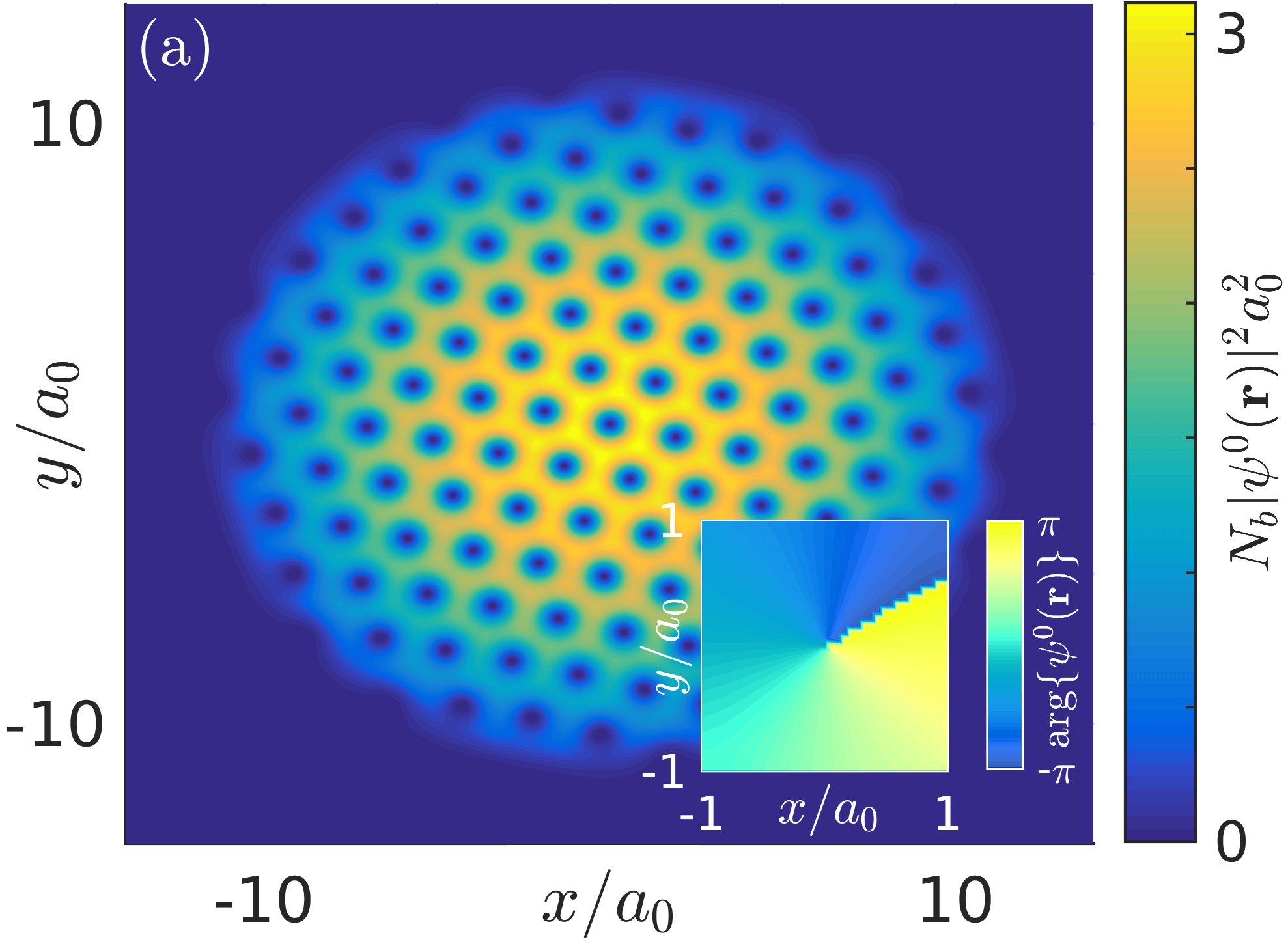}
        \end{subfigure}
        \hspace{0.15cm}
        \begin{subfigure}[c]{4.2cm}
                \includegraphics[width=\textwidth]{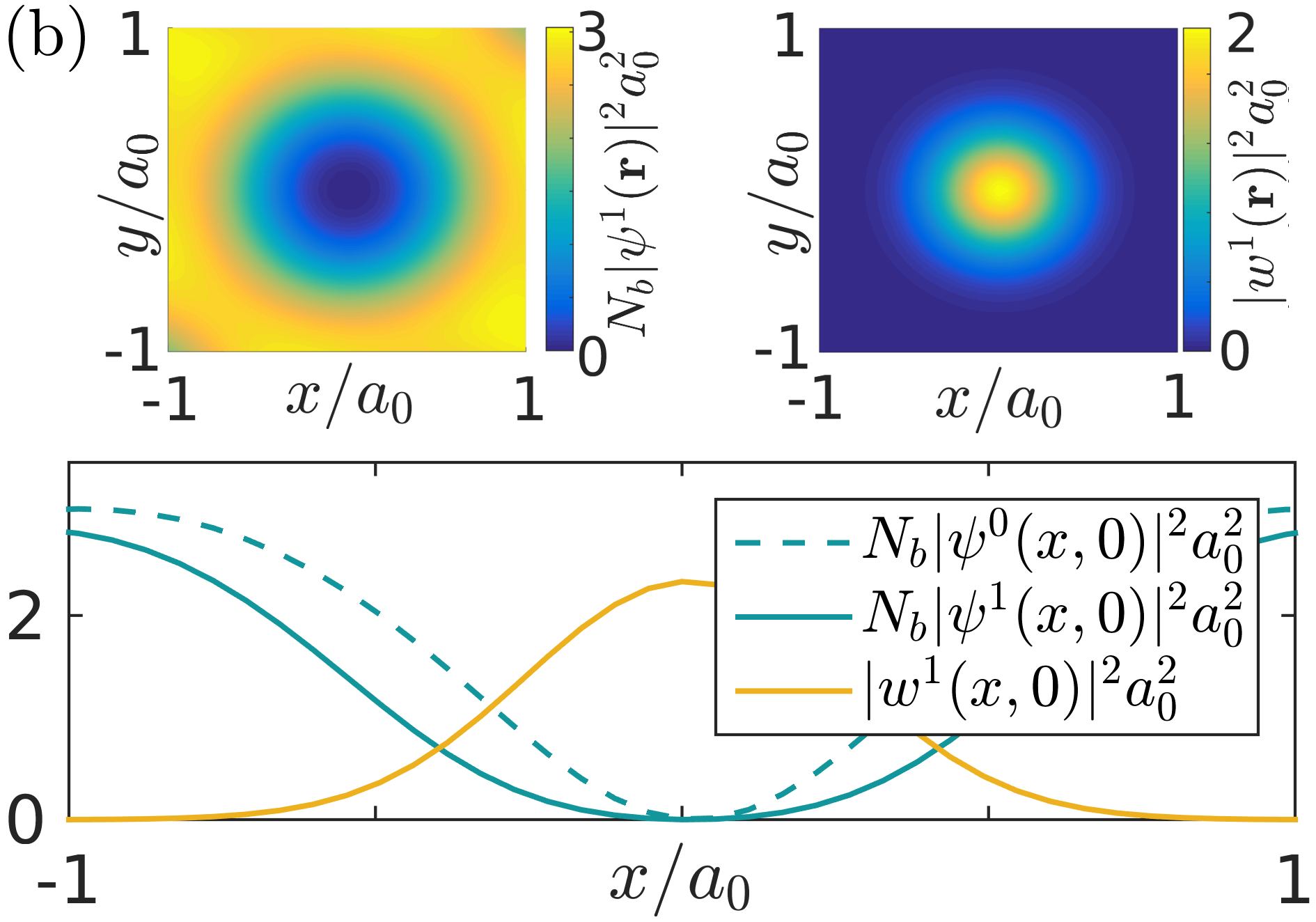}
        \end{subfigure}
        \hspace{0.15cm}
        \begin{subfigure}[c]{4.0cm}
                \includegraphics[width=\textwidth]{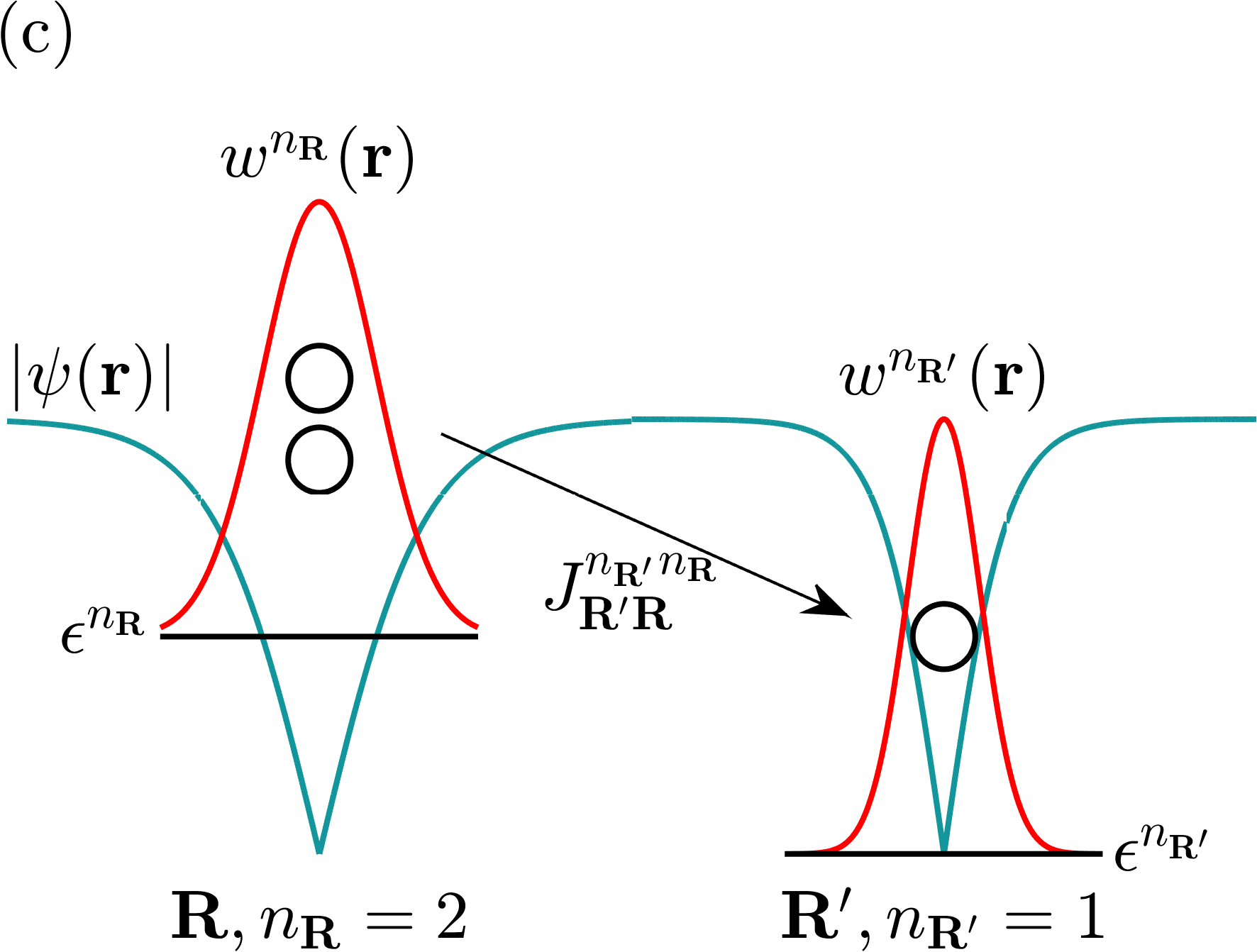}
        \end{subfigure}
        \hspace{0.15cm}
        \begin{subfigure}[c]{4.4cm}
                \includegraphics[width=\textwidth]{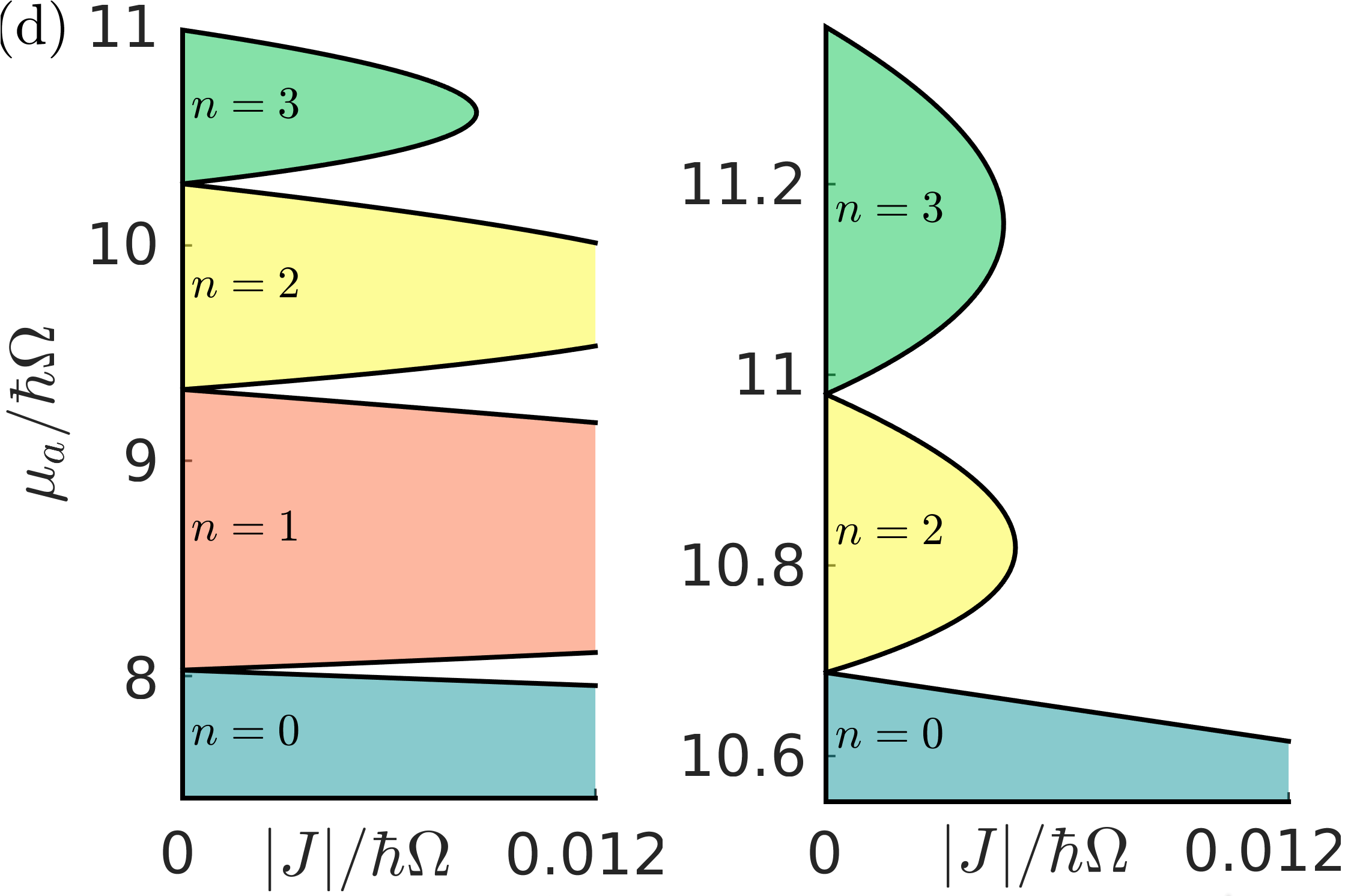}
        \end{subfigure}
        \vspace{-5pt}
\caption{\label{fig:overview} {\em Hubbard description of atoms trapped in a vortex lattice}. (a) The density $N_b |\psi^0 (\rr) |^2$ of a harmonically-trapped and rotating BEC. Inset: the phase structure of the central vortex. (b) The density of the BEC $N_b |\psi^1 (\rr) |^2$ (top left) and impurity $|w^1 (\rr) |^2$ (top right) when a single impurity is localized in the central vortex. Bottom: A cross-section of these densities along the $x$-axis, along with the unperturbed BEC density $N_b |\psi^0 (\rr) |^2$ (dashed line). The parameters for (a) and (b) are $N_b = 500$, $\Omega/\Omega_b = 0.98$, $m_b g_b / \hbar^2 = 1$, $m_b/m_a = 1$, $g_{a}/g_b = 1.1$ and $g_{ab}/g_b = 6$. Physical quantities are expressed in terms of the characteristic length $a_0 = \sqrt{\hbar/m_b \Omega_b}$ of the trap. (c) An illustration of the Hubbard physics. Impurities at a vortex site $\RR$ are described by a wavefunction $w_\RR^{n_\RR}(\rr)$ that, along with the deformation of absolute value $|\psi(\rr)|$ of the BEC wavefunction in the vicinity of $\RR$, is dependent on occupation $n_\RR$. This leads to an occupation-dependent energy $\epsilon^{n_\RR}$ per impurity at the site and occupation-dependent hopping energy $J_{\RR' \RR}^{n_{\RR'} n_{\RR}}$ between sites. (d) The Mott lobes (shaded regions) of the Hubbard phase diagram in terms of the chemical potential $\mu_a$ and hopping energy $|J|$ of the impurities, found within the Gutzwiller ansatz~\cite{Umucalhlar2007}. If $\epsilon^{n}$ increases monotonically then all Mott lobes are present, otherwise some are missing. Here we show one example for each case, corresponding to $g_{a}/g_b = 1.5$ and $g_{ab}/g_b = 3$ (left), and $g_{a}/g_b = 1.1$ and $g_{ab}/g_b = 6$ (right). The other parameters are $N_b = 10$, $m_b g_b / \hbar^2 = 1$ and $m_b/m_a = 1$.
}
\end{figure*}


{\em System.---}Our system consists of a cold atomic mixture of two bosonic species $s =a,b$, which we call impurities and bosons, respectively. They have mass $m_s$ and rotate with frequency $\Omega$ around the $z$-axis. Both species are trapped sufficiently strongly along the $z$-axis that the system is governed by a two-dimensional Hamiltonian $\hat{H} = \sum_{s = a,b} \hat{H}_s + \hat{H}_{ab}$. For each species, working in the rotating frame, we have~\cite{Fetter2009}
\begin{align}
\hat{H}_s = \int \dd \rr \cre{\Psi}_s (\rr)\left ( h_s (\rr) + \frac{g_s}{2} \cre{\Psi}_s(\rr) \an{\Psi}_s(\rr) \right ) \an{\Psi}_s(\rr), \nonumber
\end{align}
with single-particle Hamiltonian
\begin{align}
h_s (\rr) = -\frac{\hbar^2 \nabla^2}{2m_s} + V_s (\rr) + \Omega L_z (\rr) , \nonumber
\end{align}
creation $\cre{\Psi}_s (\rr)$ and annihilation $\an{\Psi}_s(\rr)$ field operators, differential operator $L_z (\rr) = - \ii \hbar \partial / \partial \phi$ for the angular momentum around the $z$-axis, and position vector $\rr$ orthogonal to the $z$-axis. The repulsive interaction between atoms is of the density-density type with intra- and inter-species interactions having strengths $g_s$ and $g_{ab}$, respectively. Accordingly, the inter-species interaction Hamiltonian is
\begin{align}
\hat{H}_{ab} = g_{ab} \int \dd \rr \cre{\Psi}_a (\rr) \an{\Psi}_a(\rr) \cre{\Psi}_b(\rr) \an{\Psi}_b(\rr) . \nonumber
\end{align}
The relationship between the parameters of this effective two-dimensional system and those of the original three-dimensional system are given in Sec.\ I of the Supplemental Material~\cite{SM}.

For isotropic harmonic potentials $V_s (\rr) = m_s \Omega^2_s r^2/2$, the single-particle Hamiltonians can be rewritten
\begin{align}
h_s (\rr) = \frac{\Pi_s^2}{2m_s} + \frac{m_s}{2} (\Omega_s^2 - \Omega^2) r^2 , \nonumber
\end{align}
with covariant momenta $\Pii_s = -\ii \hbar \nablaa + m_s \AAA (\rr)$ and vector potential $\AAA = -\Omega \rr \times \hat{\zz}$ (symmetric gauge). We choose $\Omega \lesssim \Omega_s$, ensuring the system is trapped but nevertheless approximately homogeneous $h_s (\rr) \approx \Pi_s^2 /2m_s$ in the bulk.

{\em Vortex lattice.---}We use a Gross-Pitaevskii (GP) mean-field treatment in which, at low temperatures, the $N_b$ bosons form a BEC described by wavefunction $\psi^0(\rr)$. The result, found using the normalized gradient flow method via the backward Euler Fourier pseudospectral discretization~\cite{Bao2004,Bao2006,Bao2013}, is shown in \fir{fig:overview}(a). We consider the regime of large $\Omega$, in which the condensate exhibits singly quantized vortices whose centers $\RR$ form an equilateral triangular lattice with nearest-neighbor distance $a = ( 2 \pi \hbar /\sqrt{3} m_b \Omega)^{1/2}$~\cite{Fetter2009}. In the bulk of the condensate, the density $n^0 (\rr) = N_b |\psi^0(\rr)|^2$ of the bosons provides the impurities with a periodic potential $V^0_{ab} (\rr) = g_{ab} n^0 (\rr)$ with the same geometry as the vortex lattice. The vortex cores have a width of the order of the healing length $\xi = \hbar /\sqrt{g_b n_0 m_b}$ and depth $g_{ab} n_0$, where $n_0$ is the bulk density of bosons away from the vortex cores. Unlike a typical optical lattice potential, the width $\xi$, depth $g_{ab} n_0$ and separation $a$ of potential wells can be controlled separately and are not limited by optical wavelengths. We consider large rotation energies on the order of interaction energies $\hbar \Omega \lesssim n_0 g_b / 2$, so that the widths of the wells are on the same order as their separations $\xi \lesssim a$, and bosonic densities of roughly $10$ bosons per vortex, large enough that the mean-field GP description holds~\cite{Cooper2001,Baym2005}.

{\em Hubbard physics.---}The impurities, a minority species $N_a < N_b$, are immersed in the vortex lattice. For large enough $g_{ab}$ or $m_a$, the impurities occupy bound localized states inside the potential wells of the vortex lattice. An example of an impurity localized at the central vortex is shown in \fir{fig:overview}(b). A simple calculation for a Gaussian impurity in a finite circular well of width $\xi$ and depth $g_{ab} n_0$ gives an approximate condition $g_{ab}/g_b > m_b / 2 m_a$ for localization.

It follows that the low-energy dynamics of impurities consists of hopping between the bound states at vortex lattice sites, with many-body effects accounted for by the additional energy cost of bound impurities sharing the same site. A minimal physical description of such a system is illustrated in \fir{fig:overview}(c) and corresponds mathematically to a single-band Hubbard model with occupation-dependent parameters~\cite{Dutta2014}
\begin{equation}
\label{eq:Hubbard}
\hat{H}_\mathrm{Hubbard} = E^0 + \sum_{\RR} \epsilon^{\an{n}_\RR} \an{n}_\RR  +\sum_{\langle \RR' \RR \rangle} \cre{a}_{\RR'} J^{\an{n}_{\RR'} \an{n}_\RR}_{\RR' \RR} \an{a}_\RR .
\end{equation}
Here we have introduced the usual bosonic creation, annihilation and number operators $\cre{a}_{\RR}$, $\an{a}_\RR$ and $\an{n}_\RR = \cre{a}_{\RR}  \an{a}_\RR$ for each site $\RR$. We have assumed that the effects of impurities at any one site are sufficiently localized that impurities at different sites contribute to separate terms in the energy $E^0 + \sum_{\RR} \epsilon^{n_\RR} n_\RR$ i.e.\ long-distance interactions are negligible. We expand the energy per impurity $\epsilon^{n} = \epsilon^1 + U^{n} (n-1)/2$ in terms of an effective single-impurity energy $\epsilon^1$ and two-body interaction energy $U^{n}$. Dependence of the interaction energy $U^n$ on occupation $n$, often interpreted as effective three- or higher-body interactions~\cite{Johnson2009}, occurs when strong repulsive impurity-impurity interactions affect the bound states of multiple impurities at a site, as noted in Refs.~\cite{Li2006,Johnson2009,Hazzard2010,Buchler2010,Buchler2010e,Dutta2011,Bissbort2012,Luhmann2012,Lacki2013,Major2014}. Similar reasoning, applied to the hopping of particles, leads us to restrict hopping to between nearest neighbors, denoted by the angled brackets in \eqr{eq:Hubbard}, and implies that the hopping energy $J^{n_{\RR'} n_\RR}_{\RR, \RR'}$ depends on the occupations of the sites involved.

The Hubbard model must also account for the deformation of the BEC due to interactions with the impurities~\cite{Luhmann2008,Lutchyn2009,Mering2011,Jurgensen2012}. This leads to self-trapping, in which an impurity widens the vortex in which it is localized, so increasing the attractiveness of the potential well for itself, lowering $\epsilon^1$, and  for other impurities, providing an effective negative contribution to the occupation-dependent interaction energy $U^n$~\cite{Bruderer2008}. Here we assume that the hopping $J^{n_{\RR'} n_\RR}_{\RR' \RR}$ is slow, allowing a simple treatment in which deformations of the BEC follow the impurities instantaneously. It follows that, for each possible configuration $\sigma = \{ n_\RR \}$ of the impurity occupations $n_\RR$ of the vortex cores, there is a single possible low-energy state $\ket{\sigma}$ of the system. Together these span the system's low-energy Hilbert space. Equation (\ref{eq:Hubbard}), which governs the dynamics in this low-energy space, then describes polarons, quasi-particles comprising impurities and the associated vortex lattice deformations~\cite{Bruderer2007,Klein2007,Schirotzek2009}. 

We construct the states $\ket{\sigma}$ in two steps. We first write it as a product $\ket{\sigma} = \ket{\sigma; a}\ket{\sigma; b}$, where the state $\ket{\sigma; a}$ of the impurities is approximated by a symmetrized product of $n_\RR$ impurities occupying a, potentially occupation-dependent, wavefunction $w^{n_\RR}_\RR(\rr)$ centred at each site $\RR$. Then the corresponding bosonic state $\ket{\sigma; b}$ is taken to be the ground state of the reduced bosonic Hamiltonian $\an{H}_b (\sigma) = \brackets{\sigma; a}{\hat{H}}{\sigma; a}$. The appropriate wavefunctions $w^{n_\RR}_\RR(\rr)$, and thus $\ket{\sigma; a}$, are found self-consistently with the deformed BEC, and thus $\ket{\sigma; b}$, by minimizing the energy of the system. The parameters $\epsilon^{n}$ and $J^{n_{\RR'} n_\RR}_{\RR' \RR}$ of the Hubbard Hamiltonian $\hat{H}_\mathrm{Hubbard}$ appearing in \eqr{eq:Hubbard} are then chosen such that they reproduce the action of the original Hamiltonian in this low-energy subspace $\brackets{\sigma'}{\hat{H}_{\mathrm{Hubbard}}}{\sigma} = \brackets{\sigma'}{\hat{H}}{\sigma}$. For calculation details, see Secs.\ II and III  of the Supplemental Material~\cite{SM}.


\begin{figure*}[tb]
        \begin{subfigure}[b]{2 in}
                \includegraphics[width=\textwidth]{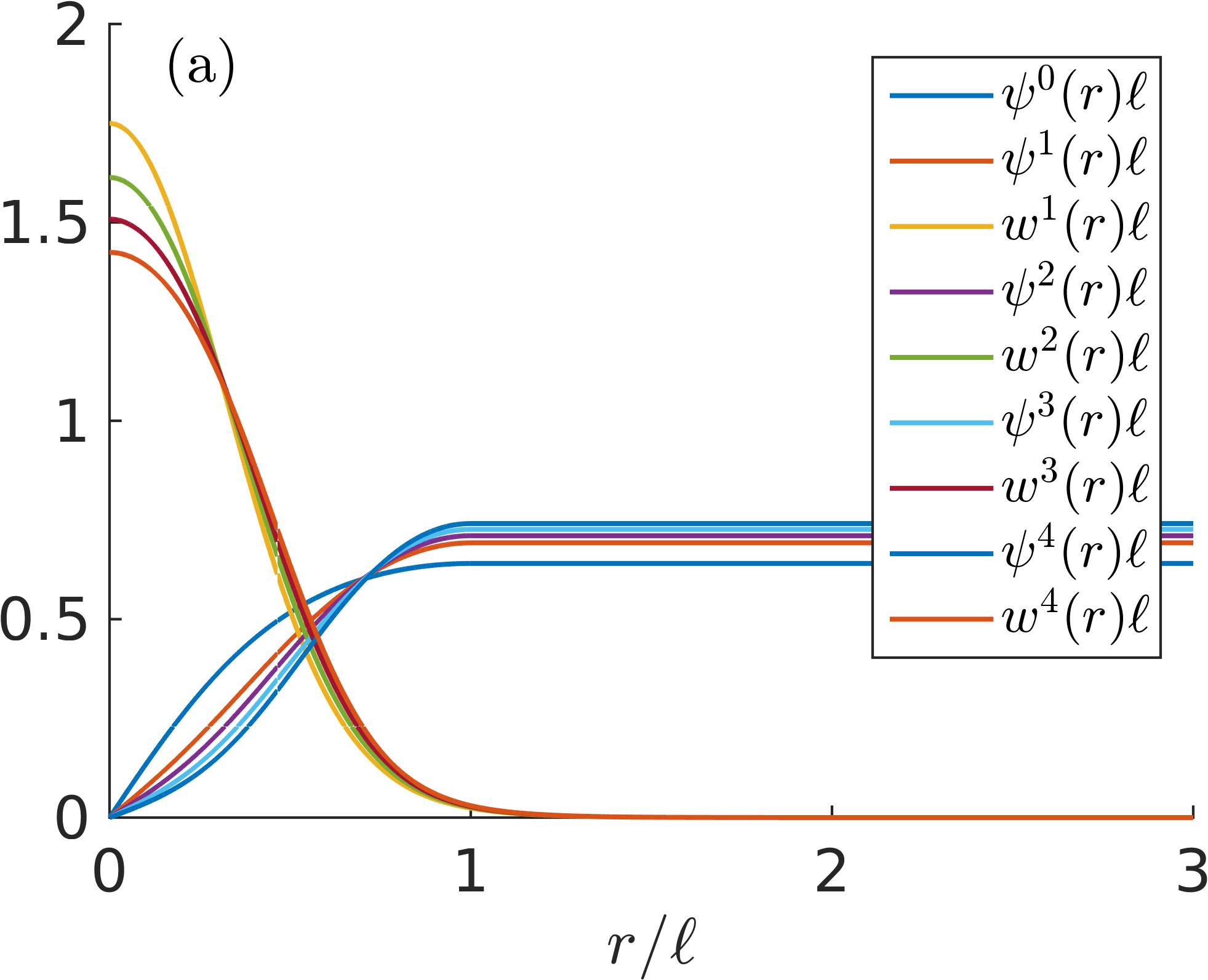}
                \label{fig:wfvsNa}
        \end{subfigure}
        \hspace{10pt}
        \begin{subfigure}[b]{2 in}
                \includegraphics[width=\textwidth]{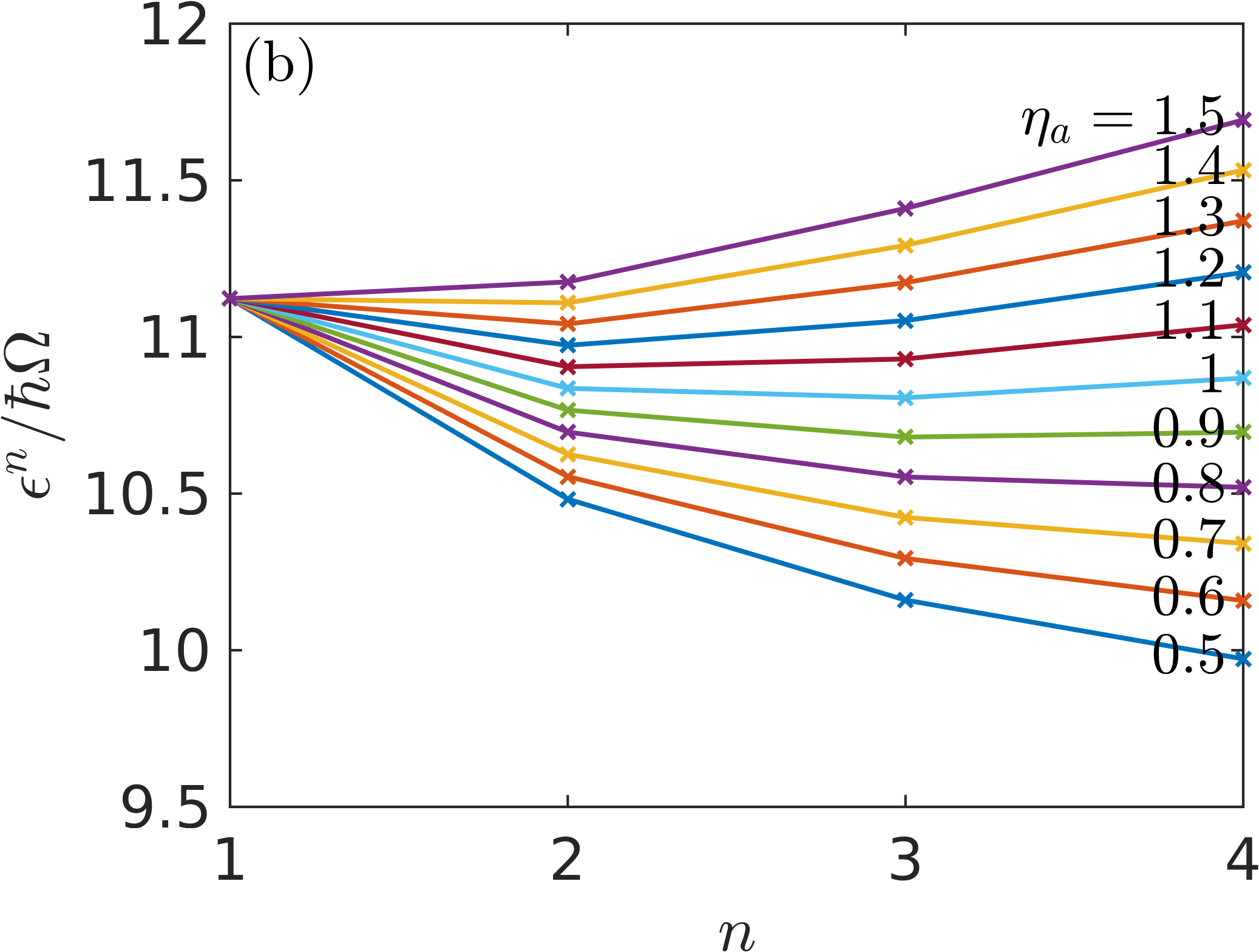}
                \label{fig:energyvsetaa}
        \end{subfigure}
        \hspace{10pt}
        \begin{subfigure}[b]{2 in}
                \includegraphics[width=\textwidth]{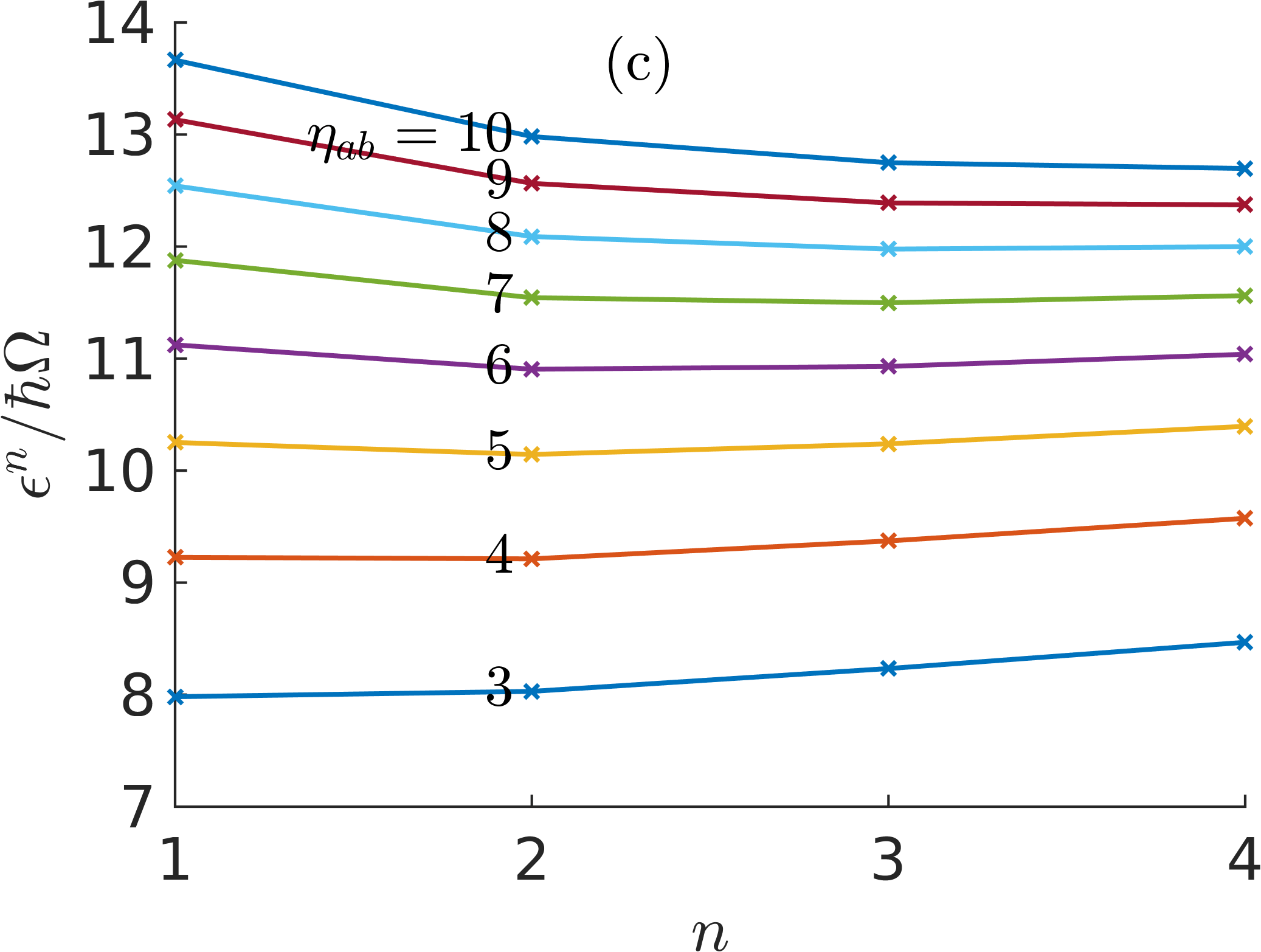}
                \label{fig:energyvsetab}
        \end{subfigure}
        \vspace{-15pt}
        \caption{\label{fig:singlecoreplots} {\em Strong coupling}. (a) The absolute value $\psi^n (r)$ of the BEC wavefunction, normalized over the unit cell $0\leq r \leq \ell$, and the impurity wavefunction $w^n (r)$, normalized over all space, for different numbers of impurities $n$. The energy per impurity $\epsilon^n$, again for different $n$, as (b) $\eta_a = g_a/g_b$ and (c) $\eta_{ab} = g_{ab}/g_b$ is varied. Crosses mark data points and lines are provided to guide the eye. Unless stated otherwise, we choose $N_b = 10$ for the number of bosons in the unit cell, $m_b g_b / \hbar^2 = 1$, $m_b/m_a = 1$, $g_{a}/g_b = 1.1$ and $g_{ab}/g_b = 6$.}
\end{figure*}


{\em Weak interactions.---}We first consider the simplest limit in which, due to weak interactions, impurities and deformations of the vortex lattice do not affect each other significantly~\cite{Bruderer2007}. In this case, the correct wavefunctions for the impurities are the localized lowest-band Wannier modes $w_\RR (\rr)$ for the vortex lattice potential $V^0_{ab} (\rr)$ formed by the unperturbed condensate $\psi^0(\rr)$. Note that the Wannier modes do not depend on occupation in this limit. We then calculate $\ket{\sigma ; b}$ within a Bogoliubov approximation of the BEC. Specifically, we write $\an{\Psi}_b (\rr) = \sqrt{N_b} \psi^0 (\rr) + \delta \an{\Psi}_b (\rr)$ with deformation $\delta \an{\Psi}_b (\rr) = \sum_\kk  [ u_\kk (\rr) \an{b}_\kk + v^\ast_\kk (\rr) \cre{b}_\kk ]$ expressed in terms of bosonic operators $\cre{b}_\kk$ and $\an{b}_\kk$, and $u_\kk (\rr)$ and $v_\kk (\rr)$ satisfying the Bogoliubov-de Gennes equations for mode energies $\hbar \omega_\kk$~\cite{Ohberg1997}. Including terms up to second-order in $\delta \an{\Psi}_b (\rr)$ and $g_{ab}$, and neglecting hopping induced by impurity-impurity interactions, we obtain the reduced bosonic Hamiltonian 
\begin{align}
\label{eq:BogHamiltonian}
\an{H}_b (\sigma) = E_b [\psi^0] + \sum_\kk \hbar \omega_\kk \cre{b}_\kk \an{b}_\kk + g_{ab} \sum_\kk \left ( f_\kk  \an{b}_\kk + f^\ast_\kk   \cre{b}_\kk \right) , \nonumber
\end{align}
with $f_\kk   = \sum_\RR n_\RR f_{\kk\RR}$, $f_{\kk\RR} = \int \dd \rr |w_\RR (\rr) |^2 f_\kk (\rr)$ and $f_\kk (\rr) = \sqrt{N_b} \left [ u_\kk (\rr) \psi^{0 \ast} (\rr) + v_\kk (\rr) \psi^0 (\rr) \right ]$. The constant $E_b [\psi^0]$ is the energy of the unperturbed condensate. Due to the simple form of $\an{H}_b (\sigma)$, its ground state is the displaced phonon vacuum
$
\ket{\sigma ; b} = \prod_{\RR}  ( \cre{X}_{\RR} )^{n_{\RR}} \ket{0} ,
$
where $\cre{X}_{\RR} = \exp [\sum_\kk (\alpha^\ast_{\kk\RR} \cre{b}_\kk - \alpha_{\kk\RR} \an{b}_\kk ) ]$ is a Glauber displacement operator with $\alpha_{\kk\RR} = -  g_{ab} f_{\kk\RR}/ \hbar \omega_\kk $.

We then find (see Sec.\ II of the Supplemental Material~\cite{SM}) the parameters of the Hubbard Hamiltonian $\hat{H}_\mathrm{Hubbard}$ [\eqr{eq:Hubbard}] to be the following. The zero energy is $E^0 = E_b[\psi^0]$, while the on-site energy per impurity $\epsilon^1 = \varepsilon - V$ consists of the contribution $\varepsilon = \int \dd \rr w^\ast_{\RR} (\rr) [ h_a (\rr) + V^0_{ab}(\rr) ]  w_{\RR} (\rr)$ from an impurity in the unperturbed potential, reduced by an energy $V = g_{ab}^2 \sum_\kk | \int \dd \rr |w_{\RR} (\rr)  |^2 f_\kk (\rr) |^2 / \hbar \omega_\kk$ due to self-trapping via the BEC deformation. The two-particle interaction $U = u - 2 V$ consists also of the usual contribution $u = g_{a}  \int \dd \rr | w_{\RR} (\rr) |^4$ for the impurity reduced by twice the self-trapping energy $V$. The hopping $J_{\RR' \RR} = \bra{0} \an{X}_{\RR'} \cre{X}_{\RR} \ket{0} j_{\RR' \RR}$ is reduced from the bare impurity hopping $j_{\RR' \RR} = \int \dd \rr w^\ast_{\RR'} (\rr) [ h_a (\rr) + V^0_{ab}(\rr) ]  w_{\RR} (\rr)$  by a renormalization factor $\bra{0} \an{X}_{\RR'} \cre{X}_{\RR} \ket{0}$ resulting from the deformation that follows each impurity. The polaron creation and annihilation operators are simply expressed as $\cre{a}_{\RR} = \left( \int \dd \rr w_\RR (\rr) \cre{\Psi}_a (\rr) \right) \cre{X}_{\RR}$ and its conjugate i.e.\ a polaron consists of an impurity with Wannier function $w_\RR (\rr)$ dressed with a corresponding displacement $\cre{X}_{\RR}$ of the BEC~\cite{Bruderer2007}.

This calculation finds the competing contributions of both the direct interaction $u$ and self-trapping $V$ to be independent of occupation. Thus the energy per impurity $\epsilon^n$ exhibits only two possible behaviors: monotonically decreasing with $n$ (attractive $U<0$), in which no Mott phases exist, or monotonically increasing with $n$ (repulsive $U>0$), leading to the usual Mott lobe structure of the phase diagram, as on the left of \fir{fig:overview}(d). Treating the deformation of the BEC in a Thomas-Fermi approximation~\cite{Johnson2012}, we find the approximate condition $g_{ab} > \sqrt{g_a g_b/2}$ for the onset of negative $U$. 

{\em Strong interactions.---}In this regime, the state of multiple impurities localized at the same site can no longer be described in terms of the lowest-band Wannier functions $w_{\RR} (\rr)$. To account for the effect of higher bands in our single-band model, we thus require occupation-dependent wavefunctions $w^{n_\RR}_{\RR} (\rr)$ to describe the impurities at each site $\RR$~\cite{Chiofalo2000,Vignolo2003,Schaff2010,Dutta2011}. Similarly, to describe the condensate in the vicinity of the site, we use an occupation-dependent wavefunction $\psi^{n_\RR}_{\RR} (\rr)$. The self-consistent wavefunctions $w^{n_\RR}_{\RR} (\rr)$ and $\psi^{n_\RR}_{\RR} (\rr)$ that optimally describe the low-energy manifold are those that simultaneously minimize the energy at the site. Omitting the site label, the energy functional to minimize is (see Sec.\ III of the Supplemental Material~\cite{SM})
\begin{equation}
\label{eq:energy}
\begin{aligned}
& E^0 + \epsilon^{n} n =\\
& n \int \dd \rr \left( w^{n} (\rr) \right)^\ast  h_a (\rr) w^{n} (\rr) + \frac{g_a}{2} n (n - 1) \int \dd \rr | w^{n} (\rr) |^4 \\
& + N_b \int \dd \rr \left( \psi^{n} (\rr) \right)^\ast h_b (\rr) \psi^{n} (\rr) + \frac{g_b}{2} N_b^2 \int \dd \rr | \psi^{n} (\rr) |^4 \\
& + n N_b g_{ab} \int \dd \rr | w^{n} (\rr)|^2 | \psi^{n} (\rr) |^2 ,
\end{aligned} \nonumber
\end{equation}
for each $n$, with $E^0$ and $\epsilon^n$ determined from the energies at the minima. 

We perform the energy minimization over the unit cell corresponding to some site $\RR$ in the bulk, approximating it by a circular region of radius $\ell = \sqrt{\hbar/m_b \Omega}$ and area equal to that of the unit cell (see Sec.\ IV of the Supplemental Material~\cite{SM}). We fix the number $N_b$ of bosons in a unit cell and use a Crank-Nicolson finite-difference approach~\cite{Adhikari2000}. We account for the BEC rotation and trapping by making two assumptions. First, that the BEC phase has a simple angular dependence $\psi^n (\rr) = \ee^{\ii \phi} \psi^n (r)$ of its wavefunction inside the unit cell [\fir{fig:overview}(a)]. Second, that the wavefunction $\psi^n(r)$ takes its maximum value at the boundary of the cell [\fir{fig:overview}(b)]. We also assume no angular dependence for the impurity wavefunction $w^n (\rr) = w^n (r)$.

The results are shown in \fir{fig:singlecoreplots}. In \fir{fig:singlecoreplots}(a) we observe the widening of both the impurity wavefunction and vortex core due to strong interactions. In Figs.\ \ref{fig:singlecoreplots}(b) and (c) we plot the occupation dependence of the energy per impurity $\epsilon^n$ for varying intra- and inter-species interaction strengths $g_{a}$ and $g_{ab}$. 
We see that the competing effects of direct repulsive interactions and attractive mediated interactions are separately occupation dependent. Specifically, the effect of self-trapping decreases quickly with occupation due to the reduced effect of an impurity on a vortex core that has already been widened. Direct interactions, meanwhile, remain important for larger $n$. Thus we find regimes in which $U^n$ is initially negative and then positive, and $\epsilon^n$ is non-monotonic, first decreasing then increasing. 

This corresponds to unusual behavior in the phase diagram, which we calculate within the Gutzwiller ansatz~\cite{Umucalhlar2007} for two sets of parameters, assuming the hopping to have a constant magnitude $|J|$ (see Sec.\ V of the Supplemental Material~\cite{SM}). The Mott lobes, up to $n=3$, are shown in \fir{fig:overview}(d). We see that as the  strength of impurity-BEC interactions $g_{ab}$ increases relative to impurity-impurity interactions $g_a$, the $n=1$ Mott lobe disappears completely.

{\em Hopping.---}While the chemical potential $\mu_a$ of the impurity can be controlled independently, the magnitude $|J|$ of the hopping depends on the other parameters. Here we estimate the magnitude $|J|$, demonstrating that we are in the region of the phase diagram containing the Mott lobes and validating our earlier assumption of slow hopping. We take two of the previously calculated single-impurity wavefunctions $w^1_\RR (\rr)$, located at neighboring sites $\RR$ and $\RR'$, orthogonalize them (see Sec.\ VI of the the Supplemental Material~\cite{SM}) and calculate the bare hopping using the unperturbed potential $j_{\RR' \RR} = \int \dd \rr w^{1 \ast}_{\RR'} (\rr) [ h_a (\rr) + V^0_{ab}(\rr) ]  w^1_{\RR} (\rr)$. For the parameters $N_b = 10$, $m_b g_b / \hbar^2 = 1$, $m_b/m_a = 1$ and $g_{ab}/g_b = 6$, corresponding to the missing Mott lobe, we find $|j_{\RR' \RR}| = 7.4 \times 10^{-3} \hbar \Omega$. The true magnitude $|J^{n_\RR n_{\RR'}}_{\RR' \RR}|$ of the hopping energies will be renormalized to a significantly smaller value than this due to deformations of the BEC. Hopping is thus the smallest energy scale in our system, one order of magnitude below $U^n$.

{\em Discussion.---}We have shown that it is possible to trap cold atomic impurities in the vortex lattice formed by rotating bosons of another species, described their motion by a Hubbard model, and shown that the Mott lobes of the resulting phase diagram have unusual features. Such features, including missing Mott lobes, can be observed in experiment, inferred from time-of-flight imaging~\cite{Greiner2002,Folling2005} since the lattice parameter $a$ is smaller than the wavelength of light. To confirm the feasibility of this, note that the energy scale of the system is determined by the rotation and trapping frequencies $\Omega \approx \Omega_s$ which can be on the order of $100 \; \mathrm{Hz}$ in magnetic traps and kHz in dipole traps. For example, $\Omega/2\pi = 3 \; \mathrm{kHz}$ is equivalent to $\hbar \Omega /k_B  = 150 \; \mathrm{nK}$. The temperature needs to be less than $0.2\hbar\Omega/k_B = 30 \; \mathrm{nK}$ to distinguish the Mott lobes, as is evident from \fir{fig:overview}(d), which is experimentally achievable. We note that our calculations neglect correlations between particles at the same site, which for strong interactions may lead to a significant reduction in the on-site energies~\cite{Bissbort2012} and enhance the occupation-dependent interaction effects.

As well as the strongly-interacting region of the phase diagram and the Mott lobes on which we have focused, strong interactions significantly affect other phases, including inducing a superfluid of paired impurities~\cite{Dutta2011}. The malleability of the lattice formed by a BEC naturally allows a certain amount of disorder and could lead to further localization of the impurities~\cite{Anderson1958}. While we have focused on bosonic impurities, fermions could also be considered in the same framework where novel vortex induced pairing effects are expected to arise.

\begin{acknowledgements}
THJ and DJ thank the National Research Foundation and the Ministry of Education of Singapore for support. WB thanks the Ministry of Education of Singapore for support through grant R-146-000-196-112. The research leading to these results has received funding from the European Research Council under the European Union's Seventh Framework Programme (FP7/2007-2013) Grant Agreement No.\ 319286 Q-MAC and the collaborative project QuProCS Grant Agreement No.\ 641277. We gratefully acknowledge financial support from the Oxford Martin School Programme on Bio-Inspired Quantum Technologies.
\end{acknowledgements}
 

\clearpage
\onecolumngrid

\begin{center}
\Large{{\bf Supplemental Material}} 

\large{Hubbard model for atomic impurities bound by the vortex lattice of a rotating BEC} 

T.\ H.\ Johnson, Y.\ Yuan, W.\ Bao, S.\ R.\ Clark, C.\ Foot, and D.\ Jaksch
\end{center}

In this Supplemental Material we provide additional details relating to the calculations presented in the main text. We begin in \secr{sec:2dH} by deriving the effective two-dimensional Hamiltonian for our system, expressing its parameters in terms of those of the full three-dimensional system. We then provide the details of our approach to deriving a Hubbard model for our system. We first do this for weak interactions, in \secr{sec:HubbardO}, using lowest-band Wannier functions to describe impurities and the Bogoliubov ansatz for the bosons. Then, in \secr{sec:HubbardNO}, we look at strong interactions, in which the occupation-dependent impurity and bosonic wavefunctions are found simultaneously. Section \ref{sec:Hartree} then details the numerical calculations used to obtain the Hubbard parameters, focusing on the on-site terms. We then calculate the phase diagram of the Hubbard model in \secr{sec:Gutzwiller}, for fixed occupation-independent hopping within the Gutzwiller ansatz. Finally, \secr{sec:tunnelling} provides a heuristic estimation of the size of the hopping energies in the system, demonstrating that we are in the strongly-interacting region of the phase diagram.

\section{Effective two-dimensional Hamiltonian}
\label{sec:2dH}

We study an imbalanced cold-atomic mixture of a few bosonic impurities (species $a$) and many bosons (species $b$) in a three-dimensional space spanned by vector $\rr$ and parametrized by Cartesian $(x, y, z)$ or cylindrical $(r, \phi, z)$ co-ordinates. The system is rotating at frequency $\Omega$ around the $z$-axis and is thus described by the Hamiltonian~\cite{Fetter2009}
\begin{align}
\label{eq:H}
\hat{H} = & \sum_{s = a,b} \hat{H}_s + \hat{H}_{ab} , \\
\hat{H}_s = & \int \dd \rr \cre{\Psi}_s (\rr)\left ( h_s (\rr) + \frac{g_s}{2} \cre{\Psi}_s (\rr) \an{\Psi}_s (\rr) \right ) \an{\Psi}_s(\rr), \\
\hat{H}_{ab} = & g_{ab} \int \dd \rr \cre{\Psi}_a(\rr) \an{\Psi}_a(\rr) \cre{\Psi}_b(\rr) \an{\Psi}_b(\rr), \\
h_s (\rr) = & -\frac{\hbar^2 \nabla^2}{2m_s} + V_s (\rr) + \Omega L_z .
\end{align}
Here we label the two species by $s=a,b$, each species having single-particle Hamiltonian $h_s(\rr)$, mass $m_s$, external potential $V_s (\rr)$, and bosonic field operators $\cre{\Psi}_s(\rr)$ and $\an{\Psi}_s(\rr)$. The operator for the angular momentum around the $z$-axis is $L_z = - \ii \hbar \partial / \partial \phi$. Also included are intra- and inter-species density-density interactions, with strengths $g_s$ and $g_{ab}$, respectively.

We consider the external potentials to have the separable form $V_s (\rr) = V_{s,\perp} (\rr_\perp) + V_{s,z} (z)$, where $\rr_\perp$ is a two-dimensional vector in the $xy$-plane, parametrized by Cartesian $(x,y)$ or polar $(r,\phi)$ co-ordinates. As a result of this, the single-particle eigenfunctions $\varphi_{s,q,u} (\rr) = \varphi_{s,q}(\rr_\perp) Z_{s,u} (z)$ of the system, orthonormalized as $\int \dd \rr_\perp \varphi^\ast_{s,q'}(\rr_\perp) \varphi_{s,q}(\rr_\perp) = \delta_{q',q}$ and $\int \dd z Z^\ast_{s,u'}(z) Z_{s,u}(z) = \delta_{u',u}$ and with eigenvalues $E_{s,q,u} = E_{s,\perp,q} + E_{s,z,u}$, will be separable. Focusing on the axial part that depends on the $z$ co-ordinate, solutions $Z_{s,u} (z)$ are labeled by quantum number $u$ and satisfy
\begin{align}
\left( -\frac{\hbar^2}{2m_s} \frac{\dd^2}{\dd z^2} +  V_{s,z} (z)  \right) Z_u (z) = E_{s,z,u} Z_{s,u} (z)  .
\end{align}
There is a corresponding equation for the $xy$-part with solutions $\varphi_{s,q}(\rr_\perp)$. Our assumption is that the axial trapping potential $V_{s,z} (z)$ felt by each species is so tight that there is never enough energy in the system to excite the axial part away from its lowest-energy solution. In other words, $E_{s,z,u} - E_{s,z,0}$, for $u \neq 0$, is much larger than the energy of the system. Thus the relevant part of the single-particle Hilbert space is spanned by functions $\varphi_{s,q,0} (\rr) = \varphi_{s,q}(\rr_\perp) Z_{s,0} (z)$ and we may perform the expansion
\begin{align}
\label{eq:fieldoperatorred}
\cre{\Psi}_s(\rr) = Z_{s,0} (z) \sum_q \varphi_{s,q}(\rr_\perp) \cre{a}_{s,q,0} = Z_{s,0} (z) \cre{\Psi}_{s, \perp} (\rr_\perp),
\end{align}
with little loss of accuracy. Here, in an intermediate step we have introduced bosonic mode creation operators $\cre{a}_{s,q,u}$ corresponding to eigenfunctions $\varphi_{s,q,u} (\rr)$.

Inserting \eqr{eq:fieldoperatorred} into the Hamiltonian of \eqr{eq:H} results in the reduced two-dimensional Hamiltonian
\begin{align}
\hat{H} = & \sum_{s = a,b} \hat{H}_s + \hat{H}_{ab} , \\
\hat{H}_s = & \int \dd \rr_\perp \cre{\Psi}_{s, \perp} (\rr_\perp)\left ( h_{s,\perp} (\rr_\perp) + \frac{g_{s, \perp}}{2} \cre{\Psi}_{s, \perp} (\rr_\perp) \an{\Psi}_{s, \perp} (\rr_\perp) \right ) \an{\Psi}_{s, \perp}(\rr_\perp), \\
h_{s, \perp} (\rr_\perp) = & E_{s,z,0} -\frac{\hbar^2 \nabla^2_\perp}{2m_s} + V_{s,\perp} (\rr_\perp) + \Omega L_z , \\
\hat{H}_{ab} = & g_{ab,\perp} \int \dd \rr_\perp \cre{\Psi}_{a,\perp}(\rr_\perp) \an{\Psi}_{a,\perp}(\rr_\perp) \cre{\Psi}_{b,\perp}(\rr_\perp) \an{\Psi}_{b,\perp}(\rr_\perp), \\
g_{s, \perp} = & g_s \int \dd z | Z_{s,0} (z) |^4 , \\
g_{ab, \perp} = & g_{ab} \int \dd z | Z_{a,0} (z) |^2 | Z_{b,0} (z) |^2 .
\end{align}
Here we have introduced the two-dimensional Laplacian $\nabla^2_\perp$ associated with $\rr_\perp$.

Since we deal entirely with this effectively two-dimensional system, for clarity, we omit all occurrences of $\perp$ in the subscripts from now on. We also drop the term $E_{s,z,0}$, which merely shifts the zero of energy by this amount.

\section{Deriving the Hubbard model: weak interactions and orthogonal Wannier functions}
\label{sec:HubbardO}
Our aim is to derive a Hubbard model that captures the low-energy dynamics of the impurities in a triangular vortex lattice formed by a Bose-Einstein condensate (BEC) of rotating bosons. For this to be valid we have two requirements, which we examine in the next two paragraphs.

First, the bosons must form a vortex lattice that is sufficiently large, homogeneous and stable that it is sensible to describe the impurities by a homogeneous Hubbard model of fixed lattice geometry. A triangular vortex lattice with parameter $a = (2 \pi \hbar / \sqrt{3} m_b \Omega )^{1/2}$ satisfying this requirement of stability and homogeneity is obtained for harmonically trapped bosons $V_{b} (\rr) = m_b \Omega_b^2 r^2/2$ with trapping frequency $\Omega_b$ close to the rotation frequency $\Omega$. Similarly, a trap $V_{a} (\rr) = m_a \Omega_a^2 r^2/2$ for $\Omega_a \approx \Omega$ ensures approximately homogeneous behavior of the impurities. We use $\RR$ to denote the triangular lattice vectors, located at each vortex core. From herein it will be common to refer to each point $\RR$ as both a vortex core and a lattice site.

Second, the lattice potential felt by the impurities must be deep enough that they are trapped by and localized at the vortex cores, which occurs when $g_{ab}$ or $m_a$ is sufficiently large. This leads to a slow exchange of impurities between neighboring sites only, with the bosons and other impurities adapting effectively instantaneously to any transfer of an impurity from one site to its neighbors. This is equivalent to making a Born-Oppenheimer approximation and is required if we are to describe the motion of impurities by a simple Hubbard model. It ensures that there is only one low-energy state $\ket{\sigma}$ corresponding to each possible configuration $\sigma = \{ n_\RR \} = n_{\RR} , n_{\RR'} , n_{\RR''},   \dots $ of impurities, where $n_\RR$ denotes the occupation i.e.\ the number of impurities localized at site $\RR$. Impurities thus move around synchronized with the deformation they impart on the BEC, with it being often helpful to refer to each impurity and its deformation as a single quasi-particle, a polaron.

In this work, we take the two requirements above as a premise and check that they are filled for the chosen parameters. In addition, in this section only, we make the assumption that the interactions between particles are weak enough that the states of the impurities are accurately described using the lowest-band Wannier functions $w_\RR (\rr)$ for non-interacting impurities moving in the unperturbed vortex lattice and the Bogoliubov approximation for the BEC is valid. In \secr{sec:HubbardNO} we will remove these last assumptions, considering stronger interactions between particles and capturing the corresponding effects.

\subsection{Wannier functions}
\label{sec:sifunctionsO}
We work under the assumption that weak interactions result in the impurities only having a small effect on each other and the bosons. Thus, when deciding on a set of impurity states that describes their low-energy Hilbert space, we treat the condensate like a fixed potential $V^0_{ab} (\rr) = g_{ab} n^0 (\rr)$, where $n^0 (\rr) = N_b |\psi^0 (\rr)|^2$ is the density of the unperturbed condensate wavefunction $\psi^0 (\rr)$ occupied by $N_b$ bosons. Additionally, since interactions between impurities are small, these impurity states are built from the single-particle wavefunctions associated with non-interacting impurities. 

The task of finding localized single-particle wavefunctions $w_\RR (\rr)$ that span the low-energy subspace of non-interacting impurities is well understood~\cite{Walters2013}. First, consider the standard situation without the terms containing $\Omega$ and $\Omega_a$ appearing in the impurity Hamiltonian, thus the single-particle Hamiltonian is $h_a (\rr)= - \hbar^2 \nabla^2/2m_a + V^0_{ab} (\rr)$. Wannier functions $w_{\RR p} (\rr)$ for band $p$ are defined as the Fourier transform
\begin{equation}\label{eq:wannier_generalized}
w_{\RR p} (\rr) = \frac{\Upsilon}{(2\pi)^{2}} \int_{\mathrm{BZ}} \dd \kk \; \ee^{-\ii \kk \cdot \RR} \ee^{\ii \phi_{\kk p} }  \varphi_{\kk p} (\rr) ,
\end{equation}
of the Bloch functions $\varphi_{\kk p} (\mathbf{r})$, which are the $p$-th band eigenfunctions of $h_a (\rr)$. Here, $\Upsilon = \sqrt{3} a^2/2$ is the area of the lattice unit cell. The wave-vectors $\kk$ run over the first Brillouin zone (BZ) of the reciprocal lattice.
The Wannier functions are translations of one another $w_{\RR p} (\rr -\RR) = w_{\0 p} (\rr)$ and they are naturally orthonormal to each other since they are merely a unitary transformation of the orthonormal Bloch eigenfunctions. To ensure each Wannier function is localized, the arbitrary phases $\ee^{\ii \phi_{\kk p}}$ associated with each Bloch function are chosen such that some appropriate cost function is minimized e.g.\ the standard deviation of the position. Such a choice of Wannier functions is often referred to as maximally localized, though here we just refer to them simply as Wannier functions~\cite{Walters2013}.

This can be extended to the case in which the system is rotating at frequency $\Omega$ and harmonically trapped with trap frequency $\Omega_a$, corresponding to the replacement $h_a (\rr) \rightarrow h_a (\rr) + m_a \Omega_a^2 r^2/2 + \Omega L_z$. If $\Omega_a \approx \Omega$ then this replacement can be captured by introducing an effective magnetic vector potential $\AAA (\rr) = - \Omega \rr \times \hat{\zz}$. The effect is to transform
\begin{equation}
-\ii \hbar \nablaa \rightarrow -\ii \hbar \nablaa + m_a \AAA (\rr),
\end{equation}
in the single-particle Hamiltonian $h_a (\rr)$. Expressions such as $[h_a(\rr) + V^0_{ab}(\rr) ] \varphi (\rr) = E \varphi (\rr)$ are left invariant so long as the wavefunctions are also transformed as
\begin{equation}
\varphi (\rr) \rightarrow \exp \left [ \frac{\ii m_a }{\hbar} \int^\rr_\0 \dd \rr' \cdot \AAA (\rr')  \right ] \varphi (\rr) ,
\end{equation}
where the path of integration is fixed. The upshot is that the appropriate choice of Wannier functions in the rotating case is given by
\begin{equation}
w_{\RR p} (\rr)  \rightarrow \exp \left [ \frac{\ii m_a }{ \hbar} \int^\rr_\0 \dd \rr' \cdot \AAA (\rr') \right ] w_{\RR p} (\rr)  ,
\end{equation}
and the path of integration is chosen to ensure the orthogonality of the Wannier functions~\cite{Bhat2006,Bhat2007}. Note, however, often this new Wannier function does not need to be calculated explicitly as integrals involving the rotating single-particle Hamiltonian and Wannier functions can be re-expressed simply in terms of the non-rotating equivalents. To give some examples that we will introduce fully later [Eqs.\ (\ref{eq:U}), (\ref{eq:varepsilon}) and (\ref{eq:j})]
\begin{align}
u_p = g_a \int \dd \rr | w_{\RR p} (\rr) |^4 \rightarrow & u_p , \\
\varepsilon_{p} = \int \dd \rr w^\ast_{\RR p} (\rr) [ h_a (\rr) + V^0_{ab} (\rr) ] w_{\RR p} (\rr) \rightarrow & \varepsilon_{p} , \\
\label{eq:peierlstrans}
j_{\RR' \RR p} = \int \dd \rr w^\ast_{\RR' p} (\rr) [ h_a(\rr)+ V^0_{ab} (\rr) ] w_{\RR p} (\rr) \rightarrow & \exp \left [ - \ii \Phi_{\RR' \RR} \right ] j_{\RR' \RR p} , 
\end{align}
where we have introduced the Peierls phase $\Phi_{\RR' \RR} = \frac{m_a }{ \hbar} \int^{\RR'}_\RR \dd \rr' \cdot \AAA (\rr') $. Importantly, we find that the on-site impurity properties, e.g.\ $u_p$ and $\varepsilon_p$, which are the key parameters in our treatment, are unaffected by the rotation (other than the shape of the potential $V_{ab} (\rr)$ itself being a result of the rotation of the bosons).

The relevant part of the impurity subspace for the lowest-energy dynamics is spanned by the Wannier functions of the lowest band. The assumption here is that the energies in the system, which includes the rotation energy $\hbar \Omega$, is smaller than the gap between the lowest and first excited band. We thus focus on these functions $w_{\RR} (\rr)$, dropping the band subscript. 
The upshot of the above is that the relevant part of the impurity Hilbert space is spanned by states $\ket{\sigma; a}$, which are symmetrized (as the impurities are identical bosons) states in which $n_\RR$ impurities are in single-particle Wannier functions $w_\RR (\rr)$ localized at sites $\RR$. The same approach could be applied to fermionic impurities, but we do not do that here. We now go on to find the corresponding states $\ket{\sigma}$ of the joint impurity-BEC system assuming the impurities to be bosonic.

\subsection{Polarons: Born-Oppenheimer approximation}
\label{sec:polaron0}
Having established a set of states $\ket{\sigma; a}$ spanning the relevant part of the impurity Hilbert space, we turn to deciding the states spanning the relevant part of the Hilbert space of the joint system. We use a Born-Oppenheimer approximation, which is the assumption that whenever the impurities are in state $\ket{\sigma; a}$, the bosons will accordingly be in $\ket{\sigma; b}$, the ground state of the reduced bosonic Hamiltonian 
\begin{align}
\label{eq:redHb}
\hat{H}_b (\sigma) &= \brackets{\sigma; a}{\hat{H}}{\sigma; a} . 
\end{align}
This is valid if the system stays in the low-energy part of the full Hilbert space, spanned by $\ket{\sigma} = \ket{\sigma; a} \ket{\sigma; b}$, which is correct provided the transitions between states are slow enough. The physical picture is that as impurities move around so do the deformations they create in the BEC. In this limit, it is possible to refer to the combination of impurity and deformation as a quasi-particle called a polaron~\cite{Bruderer2007}. 

The behavior of the Hamiltonian $\an{H}$ of the system in the reduced low-energy space is captured by the elements $H_{ \sigma'  \sigma} = \brackets{\sigma' }{\an{H}}{\sigma}$ that form an effective Hamiltonian matrix $\HH$. The correct Hubbard model $\an{H}_{\mathrm{Hubbard}}$ must reproduce this Hamiltonian matrix $\brackets{\sigma' }{\an{H}_{\mathrm{Hubbard}}}{\sigma} = H_{ \sigma'  \sigma}$. Hence the Hubbard model $\an{H}_{\mathrm{Hubbard}}$ will be determined from the matrix elements $H_{ \sigma'  \sigma}$, making them central objects in our study.

\subsection{Evaluating the effective Hamiltonian matrix}
\label{sec:effhamO}
We now evaluate the effective Hamiltonian matrix elements $H_{ \sigma'  \sigma} = \brackets{\sigma' }{\an{H}}{\sigma}$. We do this in three steps, corresponding to the three subsections that follow. We first evaluate $\bra{\sigma'; a} \hat{H} \ket{\sigma; a}$. In the next subsection we calculate the ground states $\ket{\sigma; b}$ of the reduced bosonic Hamiltonians described by the diagonal terms $\hat{H}_b(\sigma) = \brackets{\sigma; a}{\hat{H}}{\sigma; a}$. Then, following this, we use knowledge of $\ket{\sigma; b}$ to calculate the full effective Hamiltonian matrix elements $H_{ \sigma'  \sigma} = \brackets{\sigma' }{\an{H}}{\sigma}$.

\subsubsection{Effective Hamiltonian matrix: impurity part}

In calculating $\bra{\sigma'; a} \hat{H} \ket{\sigma; a}$ we assume the Wannier functions are localized enough such that we may ignore any terms corresponding to interactions between different Wannier functions, interaction-induced hopping, or any transfer of more than one impurities or over separations greater than that between nearest-neighbors. The result is shown below. The diagonal elements corresponding to configurations $\sigma$ are 
\begin{equation}
\label{eq:impelementssimple}
\begin{aligned}
 \bra{\sigma; a} \hat{H} \ket{\sigma; a}  =&   \int \dd \rr \cre{\Psi}_b(\rr) \left ( h_b (\rr) + \frac{g_b}{2} \cre{\Psi}_b(\rr) \an{\Psi}_b(\rr) \right ) \an{\Psi}_b(\rr)
\\
 & + \sum_{\RR} \left [ \int \dd \rr w_{\RR}^\ast (\rr)\left( h_a (\rr) + g_{ab} \cre{\Psi}_b(\rr) \an{\Psi}_b(\rr)  \right) w_{\RR} (\rr) \right] n_{\RR}
\\
 & + \sum_{\RR} \frac{u}{2} n_{\RR} ( n_{\RR} - 1) 
,
\end{aligned}
\end{equation}
with direct impurity-impurity interaction strength
\begin{align}
\label{eq:U}
u = g_{a}  \int \dd \rr | w_{\RR} (\rr) |^4 .
\end{align}
For the elements connecting two configurations $\sigma$ and $\sigma'$ by the transfer of one impurity from $\RR$ to $\RR'$, we have
\begin{equation}
\label{eq:impelementssimple2}
 \bra{\sigma'; a} \hat{H} \ket{\sigma; a}  =  \left [ \int \dd \rr w_{\RR'}^\ast (\rr)\left( h_a (\rr) + g_{ab} \cre{\Psi}_b(\rr) \an{\Psi}_b(\rr)  \right) w_{\RR} (\rr) \right] \sqrt{n'_{\RR'} n_\RR}
,
\end{equation}
where $n'_{\RR'}$ is the occupation of site $\RR'$ in configuration $\sigma'$.
All other elements are zero.

\subsubsection{Bosonic deformation: Bogoliubov approximation}
\label{sec:bosonicpart0}
We now find the ground state $\ket{\sigma; b}$ of the reduced bosonic Hamiltonian $\hat{H}_b (\sigma) = \brackets{\sigma; a}{\hat{H}}{\sigma; a}$. The ground state will describe the low-energy deformation that results from the presence of the impurities at the vortex cores, in configuration $\sigma$.
Since the effects of interactions are assumed to be small, we work within the Bogoliubov approximation, describing small deviations of the BEC wavefunction of the $N_b$ bosons from its unperturbed value $\psi^0 (\rr)$ in the absence of impurities~\cite{Bruderer2007}.

We start by making the expansion $\an{\Psi}_b (\rr) = \sqrt{N_b} \psi^0 (\rr) + \delta \an{\Psi}_b (\rr)$ in our expression for $\an{H}_b(\sigma)$, where $\delta \an{\Psi}_b (\rr)$ describes the deformation from the unperturbed condensate $\sqrt{N_b} \psi^0 (\rr)$. We discard terms above second order in $\delta \an{\Psi}_b (\rr)$ or $g_{ab}$. Then we express the deformation $\delta \an{\Psi}_b (\rr) = \sum_\kk \left [ u_\kk (\rr) \an{b}_\kk + v^\ast_\kk (\rr) \cre{b}_\kk \right ]$ in terms of bosonic Bogoliubov mode creation and annihilation operators $\cre{b}_\kk$ and $\an{b}_\kk$. Choosing $u_\kk (\rr)$ and $v_\kk (\rr)$ that satisfy the Bogoliubov-de Gennes equations
\begin{align}
\label{eq:bogdegennes1}
h_b (\rr) u_\kk (\rr) + g_b n^0 (\rr) (2 u_\kk (\rr) - v_\kk (\rr)) &= \hbar \omega_\kk u_\kk (\rr) ,
\\
\label{eq:bogdegennes2}
h_b (\rr) v_\kk (\rr) + g_b n^0 (\rr) (2 v_\kk (\rr) - u_\kk (\rr) )&= -\hbar \omega_\kk v_\kk (\rr)  ,
\end{align}
we obtain the following simplification of the reduced bosonic Hamiltonian
\begin{equation}
\begin{aligned}
\hat{H}_b (\sigma) = \bra{\sigma; a} \hat{H} \ket{\sigma; a}  =&   E_b [\psi^0]  + \sum_\kk \hbar \omega_\kk \cre{b}_\kk \an{b}_\kk
\\
 & + \sum_{\RR} \left [ \varepsilon + g_{ab} \sum_\kk \left ( f_{\kk \RR}  \an{b}_\kk + f^\ast_{\kk \RR}   \cre{b}_\kk \right) \right] n_{\RR}
\\
 & + \sum_{\RR} \frac{u}{2} n_{\RR} ( n_{\RR} - 1) 
,
\end{aligned}
\end{equation}
with unperturbed condensate energy
\begin{align}
E_b [\psi^0] = \int \dd \rr \psi^{0 \ast} (\rr) \left [ N_b h_b (\rr) + \frac{g_b N_b^2}{2} |\psi^0 (\rr) |^2  \right ] \psi^0 (\rr)   ,
\end{align}
unrenormalized single-impurity energy
\begin{align}
\label{eq:varepsilon}
\varepsilon = \int \dd \rr w^\ast_{\RR} (\rr) [ h_a (\rr) + V^0_{ab}(\rr) ]  w_{\RR} (\rr) ,
\end{align}
and coupling matrix elements
\begin{align}
f_{\kk\RR} =& \int \dd \rr |w_\RR (\rr) |^2 f_\kk (\rr)  ,
\\
f_\kk (\rr) =& \sqrt{N_b} \left [ u_\kk (\rr) \psi^{0 \ast} (\rr) + v_\kk (\rr) \psi^0 (\rr) \right ]  .
\end{align}

The ground state of this Hamiltonian $\hat{H}_b(\sigma)$ is the displaced phonon vacuum
\begin{align}
\ket{\sigma ; b} = \prod_{\RR}  \left ( \cre{X}_{\RR} \right)^{n_{\RR}} \ket{0} , \nonumber
\end{align}
where $\cre{X}_{\RR} = \exp [\sum_\kk (\alpha^\ast_{\kk\RR}  \cre{b}_\kk - \alpha_{\kk\RR}  \an{b}_\kk ) ]$ is a Glauber displacement operator with $\alpha_{\kk\RR}  = -  g_{ab} f_{\kk\RR} / \hbar \omega_\kk $.

\subsubsection{Effective Hamiltonian matrix: bosonic part}
The full effective Hamiltonian matrix elements are then calculated using knowledge of $\ket{\sigma ; b}$ and the properties of displacement operators $\cre{X}_{\RR}$, arriving at
\begin{equation}
\label{eq:elementsBog}
\begin{aligned}
H_{ \sigma'  \sigma}  = \bra{\sigma'} \hat{H} \ket{\sigma}  =&  \delta_{\RR' \RR} \Bigg \{ E_b [\psi^0] + \sum_{\RR''} (\varepsilon - V) n_{\RR''} + \sum_{\RR''} \left( \frac{u}{2} - V \right) n_{\RR''} ( n_{\RR''} - 1) \Bigg \}
\\
&+ (1-\delta_{\RR' \RR}) \Bigg \{  r_{\RR' \RR} j_{\RR' \RR} \sqrt{n'_{\RR'} n_\RR} \Bigg \}
.
\end{aligned}
\end{equation}
This contains a renormalization of the impurity energy to the polaron energy by the addition of a factor (which also appears in the renormalization of the interaction energy) 
\begin{align}
V = g_{ab}^2 \sum_\kk \frac{| \int \dd \rr |w_{\RR} (\rr)  |^2 f_\kk (\rr) |^2 }{ \hbar \omega_\kk} ,
\end{align}
the direct impurity hopping
\begin{align}
\label{eq:j}
j_{\RR' \RR} = \int \dd \rr w^\ast_{\RR'} (\rr) [ h_a (\rr) + V^0_{ab}(\rr) ]  w_{\RR} (\rr) , 
\end{align}
and a multiplicative renormalization factor for the hopping
\begin{align}
r_{\RR' \RR} = \brakets{\sigma'; b}{\sigma; b} =  \bra{0} \an{X}_{\RR'} \cre{X}_{\RR} \ket{0} =  \prod_\kk \ee^{ - ( |\alpha_{\kk \RR'} |^2 + |\alpha_{\kk \RR}|^2 -2 \alpha_{\kk \RR'}^\ast  \alpha_{\kk \RR} )/2} .
\end{align}

We rewrite this as
\begin{align}
\label{eq:matelO}
H_{ \sigma'  \sigma}  = \bra{\sigma'} \hat{H} \ket{\sigma}  =&  \delta_{\RR' \RR} \Bigg \{ E^0 + \sum_{\RR''} \epsilon^{n_{\RR''}} n_{\RR''} \Bigg \}
+ (1-\delta_{\RR' \RR}) \Bigg \{  J_{\RR' \RR} \sqrt{n'_{\RR'} n_\RR} \Bigg \} .
\end{align}
Here we have introduced the zero-impurity energy
\begin{align}
E^0 =& E_b [\psi^0] ,
\end{align}
the renormalized energy per impurity (or polaron)
\begin{align}
\label{eq:epsilonBog}
\epsilon^n =& \varepsilon - V + (n - 1) \left( \frac{u}{2} - V \right),
\end{align}
and the renormalized impurity (or polaron) hopping
\begin{align}
\label{eq:renormalizing}
J_{\RR' \RR} = r_{\RR' \RR} j_{\RR' \RR} .
\end{align}
It is noteworthy that, in this regime of weak interactions, the energy per impurity $\epsilon^n$ [\eqr{eq:epsilonBog}] can only decrease or increase with $n$ monotonically, depending on the relative importance of direct repulsion $u$ or self-trapping $V$ via the BEC. We show later that, for stronger interactions, this is no longer the case.

\subsection{Hubbard model}
\label{sec:HubbardsubsecO}
We are now in a position to build a Hubbard model $\an{H}_{\mathrm{Hubbard}}$ to describe our system. Since our Wannier functions $w_\RR (\rr)$ are orthonormal, so are $\ket{\sigma; a}$ and $\ket{\sigma}$. We introduce ladder operators $\cre{a}_\RR$ and $\an{a}_\RR$ satisfying the usual equations
\begin{align}
\cre{a}_\RR \ket{ n_{\RR} , n_{\RR'} , n_{\RR''},   \cdots } =& \sqrt{n_\RR+1} \ket{ n_{\RR}+1 , n_{\RR'} , n_{\RR''} , \cdots } 
, \\
\an{a}_\RR \ket{ n_{\RR} , n_{\RR'} , n_{\RR''},   \cdots } =& \sqrt{n_\RR} \ket{ n_{\RR}-1 , n_{\RR'} , n_{\RR''} , \cdots } , \\
[ \an{a}_{\RR'},  \cre{a}_\RR ] =& \delta_{\RR' \RR} .
\end{align}
It is essential that the states $\ket{\sigma} = \ket{ n_{\RR} , n_{\RR'} , n_{\RR''},   \cdots } $ are orthonormal for the ladder operators to be compatible with the commutation relation.
Then it is trivial to check that we may introduce an occupation-dependent Hubbard Hamiltonian
\begin{align}
\label{eq:HubbardO}
\an{H}_{\mathrm{Hubbard}} = E^0 + \sum_{\RR} \epsilon^{\an{n}_\RR} \an{n}_\RR  +\sum_{\langle \RR' \RR \rangle} \cre{a}_{\RR'} J_{\RR' \RR} \an{a}_\RR ,
\end{align}
that represents $\an{H}$ in the sense that it reproduces the matrix elements $H_{ \sigma'  \sigma} = \brackets{\sigma' }{\an{H}}{\sigma} = \brackets{\sigma' }{\an{H}_{\mathrm{Hubbard}}}{\sigma}$ of the effective Hamiltonian $\HH$. Here $\an{n}_\RR = \cre{a}_{\RR} \an{a}_\RR$ is the number operator and the angled brackets denote a sum over nearest neighbors. 

We can think of ladder operators $\cre{a}_\RR$ and $\an{a}_\RR$ as being creation and annihilation operators of some quasi-particle consisting of an impurity localized at $\RR$ and a contribution to the deformation of the bosonic system near $\RR$. We call this quasi-particle a polaron. In this particular case, we have the simple expression $\cre{a}_\RR = \cre{X}_\RR \left( \int \dd \rr w_\RR (\rr) \cre{\Psi}_a (\rr) \right)$ for the ladder creation operator, and similarly for the annihilation operator. The composition of a polaron by an impurity in a Wannier mode and the corresponding Glauber displacement of the BEC is clear.

\section{Deriving the Hubbard model: strong interactions and non-orthogonal well functions}
\label{sec:HubbardNO}
We now consider the regime of strong interactions. In the past few years, researchers have explored the effects of strong interactions, starting with a single species in a fixed lattice potential and how these interactions affect the Hubbard model describing the system~\cite{Li2006,Johnson2009,Hazzard2010,Buchler2010,Buchler2010e,Dutta2011,Bissbort2012,Luhmann2012,Lacki2013,Dutta2014,Major2014}.
It is clear that the method of the previous \secr{sec:HubbardO} fails for strong interactions. Strong repulsive interactions between impurities will lead to the spreading out of multiple impurities at the same lattice site. This cannot possibly be described by the usual lowest-band Wannier functions. Higher bands must enter the description. Rather than include multiple bands, a common approach to achieve this is to adjust the single-band model to include the effects of interactions. The state of the impurities at a lattice site is no longer assumed to be fixed, it is allowed to depend on the number of impurities at that site, and is found by minimizing the energy associated with the site containing a given number of impurities. An improved Hilbert space for the impurities is then built from such states and the Hubbard model describes the dynamics of the impurities in this space. The striking feature of these Hubbard models is that the Hubbard parameters will depend on occupation, hence they are often referred to as occupation-dependent Hubbard models. 

Perhaps the simplest approach, which we take, is to assume that the impurities at a single site will all have the same wavefunction and then find this wavefunction by minimizing a Gross-Pitaevskii-like energy functional describing the impurities at that site~\cite{Chiofalo2000,Vignolo2003,Schaff2010,Dutta2011}. While this is a coarse approach, neglecting correlations between the impurities at the same lattice site~\cite{Bissbort2012}, it captures the main effect of occupation dependence and the spreading of impurities due to repulsion at multiply-occupied sites. We call the occupation-dependent wavefunctions obtained in such calculations the well functions, to distinguish them from Wannier functions, a term we reserve for wavefunctions obtained from single-particle Bloch functions. 

In our case, when calculating these well functions, we also need to include the effects of repulsive interactions between the impurities and the BEC, which affects the state of the impurities at a lattice site~\cite{Luhmann2008,Lutchyn2009,Mering2011,Jurgensen2012}. In contrast to repulsive impurity-impurity interactions, the effect will be to akin to an attractive interaction and will squeeze the impurities. The impurities will deform the BEC, widening the vortex, which then localizes that and other impurities further~\cite{Bruderer2008,Johnson2012}. Thus the deformation of the BEC wavefunction must also be included in the calculation of the well functions. We do this in a self consistent way. We combine the energy minimization used to calculate the well functions with the energy minimization used to calculate the local deformation of the BEC around the well functions in the Born-Oppenheimer approximation [Secs.\ \ref{sec:polaron0} and \ref{sec:bosonicpart0}].

A side-effect of allowing occupation-dependent well functions to describe the impurities is that the well functions at different sites can no longer be orthogonal. This non-orthogonality must be taken into account in our Hubbard model but we show that, to leading order, the non-orthogonality of the well functions will have no effect on the on-site terms appearing in the Hubbard model, which are our primary focus. However, non-orthogonality does affect non-local terms, such as hopping, and we provide only a heuristic description of hopping terms, leaving this until \secr{sec:tunnelling}.

\subsection{Well functions and BEC deformations: a self-consistent approach}
\label{sec:sifunctionsNO}
We use impurity states $\ket{\sigma; a}$ that consist of symmetrized states in which $n_\RR$ impurities occupy well functions $w^{n_\RR}_\RR (\rr)$ localized at site $\RR$. Note the additional dependence on occupation number $n_\RR$. As in the previous \secr{sec:HubbardO}, we use the Born-Oppenheimer approximation that whenever the impurities are in state $\ket{\sigma; a}$, the bosons will accordingly be in $\ket{\sigma; b}$, the ground state of the reduced bosonic Hamiltonian 
\begin{align}
\label{eq:redHb}
\hat{H}_b(\sigma) &= \brackets{\sigma; a}{\hat{H}}{\sigma; a} . 
\end{align}
But differently to \secr{sec:HubbardO}, the impurity states $\ket{\sigma; a}$ are now also dependent on the states of the bosons $\ket{\sigma; b}$. The well functions $w^{n_\RR}_\RR (\rr)$ are chosen to be those that minimize the energy $\hat{H}_a(\sigma) = \brackets{\sigma; b}{\hat{H}}{\sigma; b}$. This naturally leads us to simultaneously find $w^{n_\RR}_\RR (\rr)$ and the BEC deformations around each site by minimizing terms of the form of the effective Hamiltonian matrix diagonal elements $H_{ \sigma  \sigma} = \brackets{\sigma }{\an{H}}{\sigma}$ with respect to the well function and BEC deformation. 

The well function $w^{n_\RR}_\RR (\rr)$ is assumed to depend only on the occupation $n_\RR$ of the corresponding site $\RR$. Similarly, we assume that the localization of the well functions means that diagonal matrix elements have the form we found in \secr{sec:effhamO} [\eqr{eq:matelO}]
\begin{align}
H_{ \sigma  \sigma} = E^0 + \sum_\RR \epsilon^{n_\RR} n_\RR ,
\end{align}
in which different sites enter in separate terms that are summed. We then simultaneously calculate $E^0$, $\epsilon^{n_\RR}$, $w^{n_\RR}_\RR (\rr)$ and the local BEC deformation, by performing an energy minimization for the configuration $\sigma$ consisting of $n_\RR$ impurities at site $\RR$, with no other impurities. Since large interactions between the impurities and BEC lead to large deformations of the BEC, we cannot use the Bogoliubov ansatz as we did in \secr{sec:bosonicpart0} to describe the BEC deformation. Instead, we assume the BEC to consist of a large number of bosons in the same wavefunction $\psi^{n_\RR}_\RR (\rr)$, which should capture the local deformation of the BEC by the impurities at site $\RR$.

Dropping the $\RR$ subscripts for clarity for the remainder of this subsection, the above reasoning leads us to make the Hartree ansatz for the impurities and a BEC ansatz for the bosons
\begin{align}
\ket{w^{n}} = & \frac{1}{\sqrt{n!}} \left( \int \dd \rr \cre{\Psi}_a (\rr) w^{n} (\rr)  \right)^{n} \ket{0} , \\
\an{\Psi}_b (\rr) = & \sqrt{N_b} \psi^{n}  (\rr),
\end{align}
and minimize the energy functional
\begin{equation}
\label{eq:energy}
\begin{aligned}
E_0 + \epsilon^{n} n = H_{ \sigma  \sigma} = E[w^{n},\psi^{n}] = & n \int \dd \rr \left( w^{n} (\rr) \right)^\ast  h_a (\rr) w^{n} (\rr) + \frac{g_a}{2} n (n - 1) \int \dd \rr | w^{n} (\rr) |^4 \\
& + N_b \int \dd \rr \left( \psi^{n} (\rr) \right)^\ast h_b (\rr) \psi^{n} (\rr) + \frac{g_b}{2} N_b^2 \int \dd \rr | \psi^{n} (\rr) |^4 \\
& + n N_b g_{ab} \int \dd \rr | w^{n} (\rr)|^2 | \psi^{n} (\rr) |^2 ,
\end{aligned}
\end{equation}
within this ansatz, subject to normalization conditions. From the minimizing functions $w^{n} (\rr)$ and $\psi^n (\rr)$, and the minimum value $E[w^{n},\psi^{n}] = E^0 + \epsilon^{n} n$ taken by the functional, for several $n$, we can calculate $E^0$ and $\epsilon^{n}$. These results then allow us to completely determine the diagonal elements $H_{ \sigma  \sigma} = E^0 + \sum_\RR \epsilon^{n_\RR} n_\RR$ of the effective Hamiltonian matrix for an arbitrary configuration $\sigma$.

A full discussion of the details of numerical minimization and the results for impurity $w^{n} (\rr)$ and BEC $\psi^{n} (\rr)$ wavefunctions, as well as energies $\epsilon^{n}$, are reserved for \secr{sec:Hartree}. Before this, we continue with the derivation of the Hubbard model.  

\subsection{Effect of non-orthogonality on the Hubbard model}
Since the well functions $w^{n_\RR}_\RR (\rr)$ are occupation dependent, it is not possible for all functions at different sites to be simultaneously orthogonal. Thus neither $\ket{\sigma; a}$ nor $\ket{\sigma}$ will be orthogonal in general. It was the orthogonality of these basis states that allowed us to move simply, in \secr{sec:HubbardsubsecO}, from the elements $H_{ \sigma'  \sigma}$ of the effective Hamiltonian matrix $\HH$ to the Hubbard Hamiltonian $\an{H}_{\mathrm{Hubbard}}$.
To obtain a Hubbard Hamiltonian $\an{H}_{\mathrm{Hubbard}}$, we must therefore first change to an orthonormal basis $\ket{\bar{\sigma}}$ that spans the same space as $\ket{\sigma}$, and then calculate elements $\bar{H}_{ \sigma'  \sigma} = \bra{\bar{\sigma}'} \an{H} \ket{\bar{\sigma}}$ of the effective Hamiltonian matrix $\bar{\HH}$ in this new basis. Note that we are using bars to denote objects relating to the new orthonormal basis.
  
We choose states $\ket{\bar{\sigma}}$ that are not only orthonormal but retain a local description, such that configuration $\sigma = \{ n_{\RR} \} $ still corresponds approximately to $n_\RR$ impurities localized at each site $\RR$. To this end we choose the transformation
\begin{align}
\label{eq:transformation}
\ket{\bar{\sigma}'} =& \sum_{\sigma } S_{\sigma' \sigma} \ket{\sigma} ,
\end{align}
where $S_{\sigma' \sigma}$ are elements of transformation matrix $\SSS = \GG^{-1/2}$ and $\GG$ is the normalization matrix with elements,
\begin{align}
\label{eq:transformation}
 G_{\sigma' \sigma} =& \brakets{\sigma'}{\sigma} .
\end{align}
For localized well functions $w^{n_\RR}_\RR (\rr)$, this transformation will orthogonalize while retaining localization. To see this, consider the case that quantities involving overlaps of functions localized to different sites are small and we only need to expand to first order in such quantities. To this order, the transformation is given by
\begin{align}
\ket{\bar{\sigma}'} = \ket{\sigma'} - \sum_{\sigma\neq \sigma'}  \brakets{\sigma'}{\sigma} \ket{\sigma}/2, 
\end{align}
which retains the local character of the states, only mixing in enough of states corresponding to configurations differing by a nearest-neighbor transition to provide orthogonality. The diagonal elements are unchanged $\bar{H}_{ \sigma  \sigma} = H_{ \sigma  \sigma}$ to leading order, while non-zero off-diagonals $\bar{H}_{ \sigma'  \sigma} = H_{ \sigma'  \sigma} - \brakets{\sigma'}{\sigma} (H_{ \sigma'  \sigma'} + H_{ \sigma  \sigma})/2$ contain a first-order correction to account for their overestimation in a non-orthogonal basis. The Hubbard Hamiltonian $\an{H}_{\mathrm{Hubbard}}$ calculation then proceeds as in \secr{sec:HubbardsubsecO}, using the effective Hamiltonian matrix $\bar{\HH}$. Specifically, we choose $\an{H}_{\mathrm{Hubbard}}$ such that $\bra{\bar{\sigma}'} \an{H}_{\mathrm{Hubbard}} \ket{\bar{\sigma}} = \bra{\bar{\sigma}'} \an{H} \ket{\bar{\sigma}}$.

The most important thing to note is that the diagonal elements of $\bar{\HH}$ and thus on-site terms in the Hubbard Hamiltonian
\begin{align}
\label{eq:HubbardNO}
\an{H}_{\mathrm{Hubbard}} = E^0 + \sum_{\RR} \epsilon^{n_\RR} \an{n}_\RR  +\sum_{\langle \RR' \RR \rangle} \cre{a}_{\RR'} J^{\an{n}_{\RR'} \an{n}_\RR}_{\RR' \RR} \an{a}_\RR ,
\end{align}
may be calculated, to first order, directly from the non-orthogonal well-functions~\cite{Dutta2014}. Contrastingly, the off-diagonals and non-local terms, denoted by $J^{n_{\RR'} n_\RR}_{\RR' \RR}$ in \eqr{eq:HubbardNO} and which we have assumed to be non-zero only for nearest-neighboring sites, are affected by the non-orthogonality. Thus here we only account for them heuristically, as discussed in \secr{sec:tunnelling}.

\section{Numerical simulation of localized impurities in a vortex core of a rotating BEC}
\label{sec:Hartree}
The previous two sections, and the relevant part of the main text, have established how we go about deriving the Hubbard parameters for our system of impurities moving in the vortex cores of a BEC's vortex lattice. What remains is to justify the assertions, described in \secr{sec:HubbardO}, on which this approach rests: that interactions between the BEC are strong enough to localize the impurities in the vortex cores (sites) while not deforming the BEC so much that its lattice geometry is altered. We also provide details of methods and results for the numerical calculation of the well functions $w_\RR^{n_\RR} (\rr)$ describing $n_\RR$ impurities at a site $\RR$, the deformation of the vortex core due to the presence of impurities, described by $\psi_\RR^{n_\RR} (\rr)$, and the energy costs $\epsilon^{n_\RR}$ per impurity of this configuration. These calculations determine the local Hubbard parameters.

We begin in \secr{sec:vloverview} by numerically calculating a vortex lattice formed by the rotating bosons in a harmonic trap. We establish that the centre of the system displays bulk behavior associated with a homogeneous system and therefore choose $\RR$ to be the central site. 
We determine the properties of the impurities when localized there and their effect on the vortex core shape, and calculate the corresponding energy per impurity.

Following this, in \secr{sec:singlevortex}, we use our understanding of the behavior a lattice site in the bulk of the system to set up a simpler numerical analysis involving a single vortex core. We use these numerics to calculate the properties of the Hubbard model for a strongly-interacting system, introduced in \secr{sec:HubbardNO}.

The main difference between the two calculations is that, in the former (\secr{sec:vloverview}), the density of the BEC expelled from a vortex core can be deposited elsewhere in the lattice, while, in the latter (\secr{sec:singlevortex}), it must remain in the same unit cell. Effects due of this transfer of BEC density between sites are not included in our model, but we show that it does not affect our conclusions by witnessing the same main set of effects in both types of calculation.

\subsection{Vortex lattice formed by trapped bosons}
\label{sec:vloverview}

\subsubsection{Energy functional}
In \secr{sec:sifunctionsNO} we reduced the calculation of the Hubbard parameters to the minimization of the following energy functional
\begin{equation}
\label{eq:energy2}
\begin{aligned}
E[w^{n},\psi^{n}] = & n \int \dd \rr \left( w^{n} (\rr) \right)^\ast  h_a (\rr) w^{n} (\rr) + \frac{g_a}{2} n (n - 1) \int \dd \rr | w^{n} (\rr) |^4 \\
& + N_b \int \dd \rr \left( \psi^{n} (\rr) \right)^\ast h_b (\rr) \psi^{n} (\rr) + \frac{g_b}{2} N_b^2 \int \dd \rr | \psi^{n} (\rr) |^4 \\
& + n N_b g_{ab} \int \dd \rr | w^{n} (\rr)|^2 | \psi^{n} (\rr) |^2 ,
\end{aligned}
\end{equation}
with $w^{n}(\rr)$ the wavefunction of $n$ impurities at a single site, and $\psi^n (\rr)$ the wavefunction that captures the deformation of the BEC around those impurities.

Here we begin by considering a harmonic trap $V_{b} (\rr) = m_b \Omega_b^2 r^2/2$ for the bosons, symmetric about a central site. The number $n$ and $N_b$ of impurities and bosons, respectively, are fixed and thus, using the method of Lagrange multipliers, we minimize the grand-canonical energy functional
\begin{align}
E_{\mu_a, \mu_b} [w^{n},\psi^{n}] = E [w^{n},\psi^{n}] - \mu_a n \int \dd \rr  | w^{n}(\rr) |^2 - \mu_b N_b  \int \dd \rr | \psi^{n} (\rr) |^2 ,
\end{align}
with respect to $w^{n}(\rr)$ and $\psi^{n} (\rr)$ and Lagrange multipliers $\mu_a$ and $\mu_b$. The minimum must satisfy
\begin{align}
\frac{\delta E_{\mu_a, \mu_b} [w^{n},\psi^{n}] }{\delta w^{n}} = \frac{\delta E_{\mu_a, \mu_b} [w^{n},\psi^{n}] }{\delta \psi^{n}} = 0 , 
\end{align}
a condition that is equivalent to the Euler-Lagrange equations
\begin{align}
\left( h_a (\rr) + g_{ab} N_b | \psi^{n} (\rr) |^2 +  g_a (n - 1) | w^{n} (\rr) |^2 \right) w^{n} (\rr) = & \mu_a w^{n}  (\rr), \\
\left( h_b (\rr) + g_{ab} n | w^{n} (\rr) |^2 +  g_b N_b | \psi^{n} (\rr) |^2 \right) \psi^{n} (\rr) = & \mu_b \psi^{n} (\rr) . 
\end{align}
These are a pair of coupled Gross-Pitaevskii equations.

\subsubsection{Characteristic quantities}

To get them in a form appropriate for solving numerically, we rewrite the equations in terms of dimensionless quantities, using the characteristic length $a_0 = \sqrt{\hbar/m_b \Omega_b}$ and energy $\hbar \Omega_b$ scales of the harmonic trap for species $b$. Denoting the dimensionless quantities by a prime, we use $\rr = \rr' a_0$, $w^{n} (\rr) = w^{n \prime} (\rr') / a_0$, $E = E' \hbar \Omega_b$, $\Omega_b = \Omega_b' \Omega_b$ (i.e.\ $\Omega'_b = 1$), and similarly for other quantities of the same dimensions. This results in grand-canonical energy functional
\begin{align}
E'_{\mu'_a, \mu'_b} [w^{n \prime},\psi^{n \prime}] = E'[w^{n \prime},\psi^{n \prime}] - \mu'_a n  \int \dd \rr'  | w^{n \prime} (\rr') |^2  - \mu'_b N_b \int \dd \rr' | \psi^{n \prime} (\rr') |^2 ,
\end{align}
with energy functional
\begin{equation}
\begin{aligned}
E'[w^{n \prime},\psi^{n \prime}] = & n \int \dd \rr' \left( w^{ n \prime} (\rr') \right)^\ast h'_a (\rr') w^{n \prime} (\rr') + \frac{\eta_a \gamma}{2} n (n - 1) \int \dd \rr' | \chi' (\rr') |^4 \\
& + N_b \int \dd \rr' \left( \psi^{n \prime} (\rr') \right)^\ast h'_b (\rr') \psi^{n \prime} (\rr') + \frac{\gamma}{2} N_b^2 \int \dd \rr' | \psi^{n \prime} (\rr') |^4 \\
& + n N_b \eta_{ab} \gamma \int \dd \rr' | w^{n \prime} (\rr') |^2 | \psi^{n \prime} (\rr') |^2 ,
\end{aligned}
\end{equation}
single-particle Hamiltonians
\begin{align}
h'_a (\rr') = & -\frac{\alpha \nabla^{\prime 2}}{2} + V'_a (\rr') + \Omega' L'_{z} , \\
h'_b (\rr') = & -\frac{\nabla^{\prime 2}}{2} + \frac{1}{2} r^{\prime 2} + \Omega' L'_{z} ,
\end{align}
and coupled Gross-Pitaevskii equations
\begin{align}
\left( h'_a (\rr') + \eta_{ab} \gamma N_b | \psi^{n \prime} (\rr') |^2 +  \eta_a \gamma (n - 1) | w^{n \prime} (\rr') |^2 \right) w^{n \prime} (\rr') = & \mu'_a w^{n \prime} (\rr'), \\
\label{eq:lastbeforedefs}
\left( h'_b (\rr') + \eta_{ab} \gamma n | w^{n \prime} (\rr') |^2 +  \gamma N_b | \psi^{n \prime} (\rr') |^2 \right) \psi^{n \prime} (\rr') = & \mu'_b \psi^{n \prime} (\rr') .
\end{align}
Here, the differential operators $\nabla^{\prime 2}$ and $L'_{z}$ act with respect to dimensionless quantities.

We thus see it is natural to introduce dimensionless quantities $\eta_a = g_a/g_b$ and $\eta_{ab} = g_{ab}/g_b$ for the relative interaction strengths, $\alpha = m_b/m_a$ for the mass ratio, $\gamma = m_b g_b / \hbar^2$ as the quantity that signifies the importance of interaction energies relative to kinetic energies, and the relative rotation frequency $\Omega' = \Omega /\Omega_b$. These quantities, together with $n$ and $N_b$, entirely characterize the physical properties of the system.

\subsubsection{Bosons forming a vortex lattice}
\label{sec:vl}


\begin{figure}[tb]
        \begin{subfigure}[b]{2.2 in}
                \includegraphics[width=\textwidth]{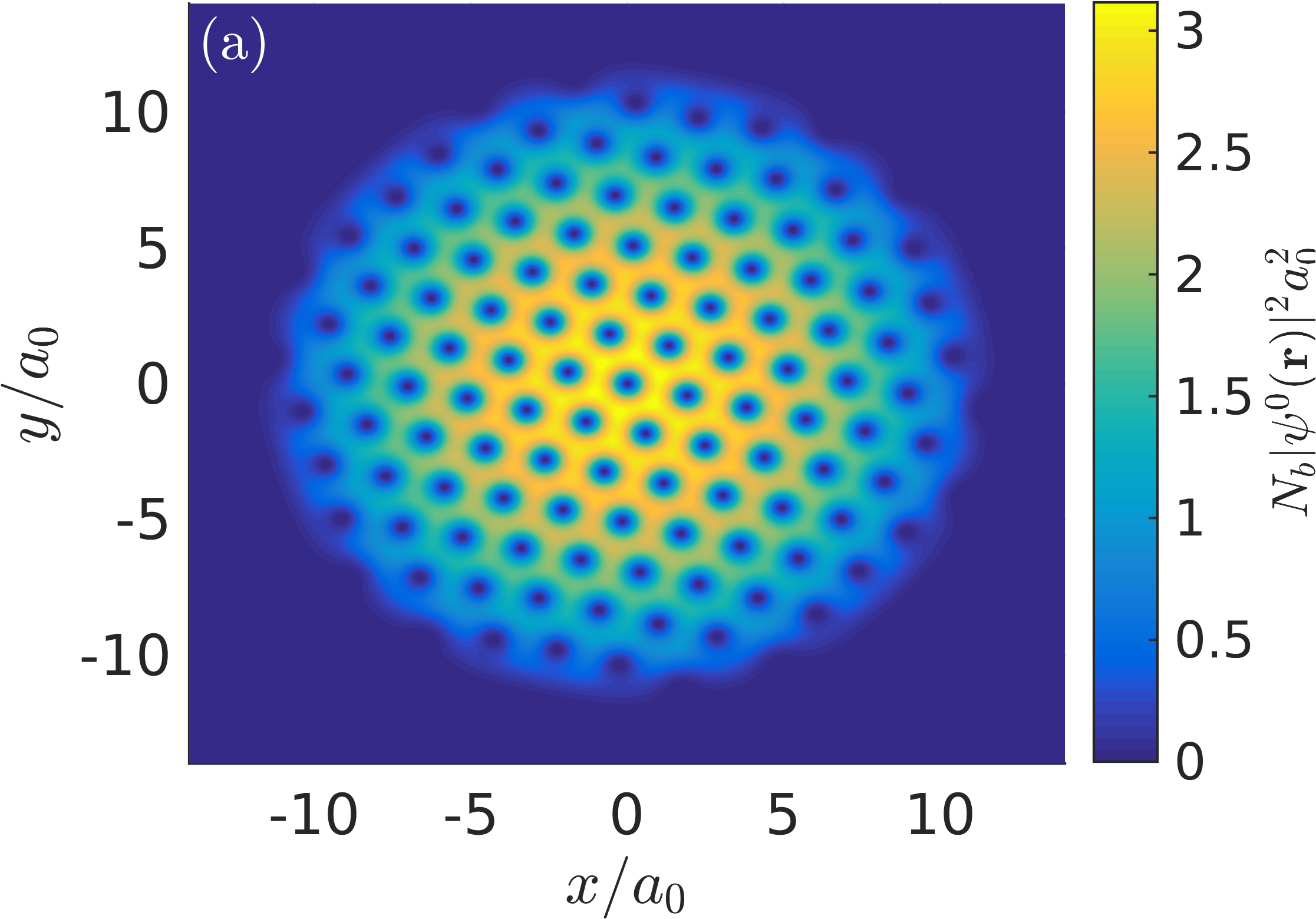}
                \label{fig:becdens0}
        \end{subfigure}
        \hspace{7pt}
        \begin{subfigure}[b]{2.2 in}
                \includegraphics[width=\textwidth]{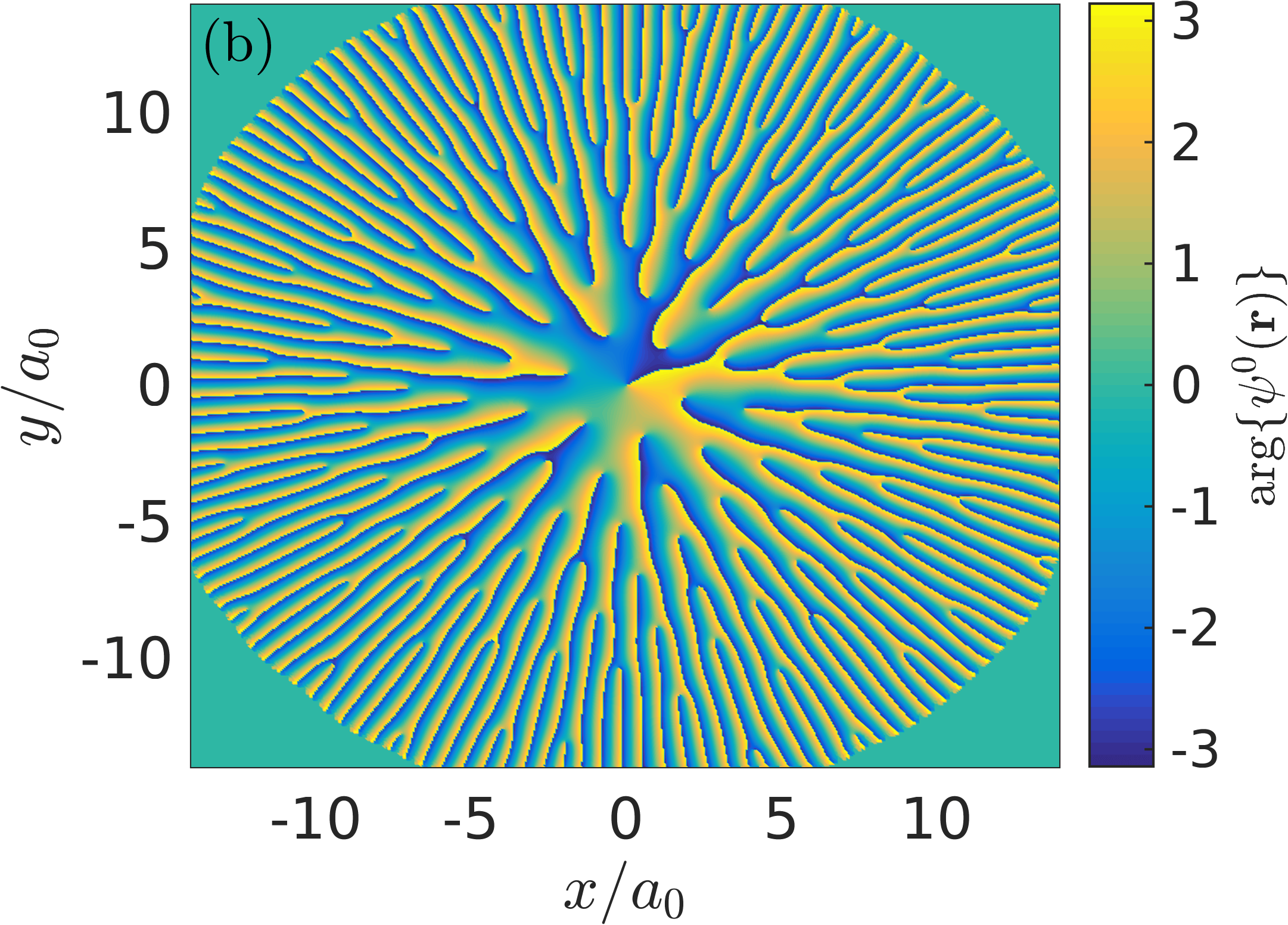}
                \label{fig:becphase0}
        \end{subfigure}
        
        \begin{subfigure}[b]{2.2 in}
                \includegraphics[width=\textwidth]{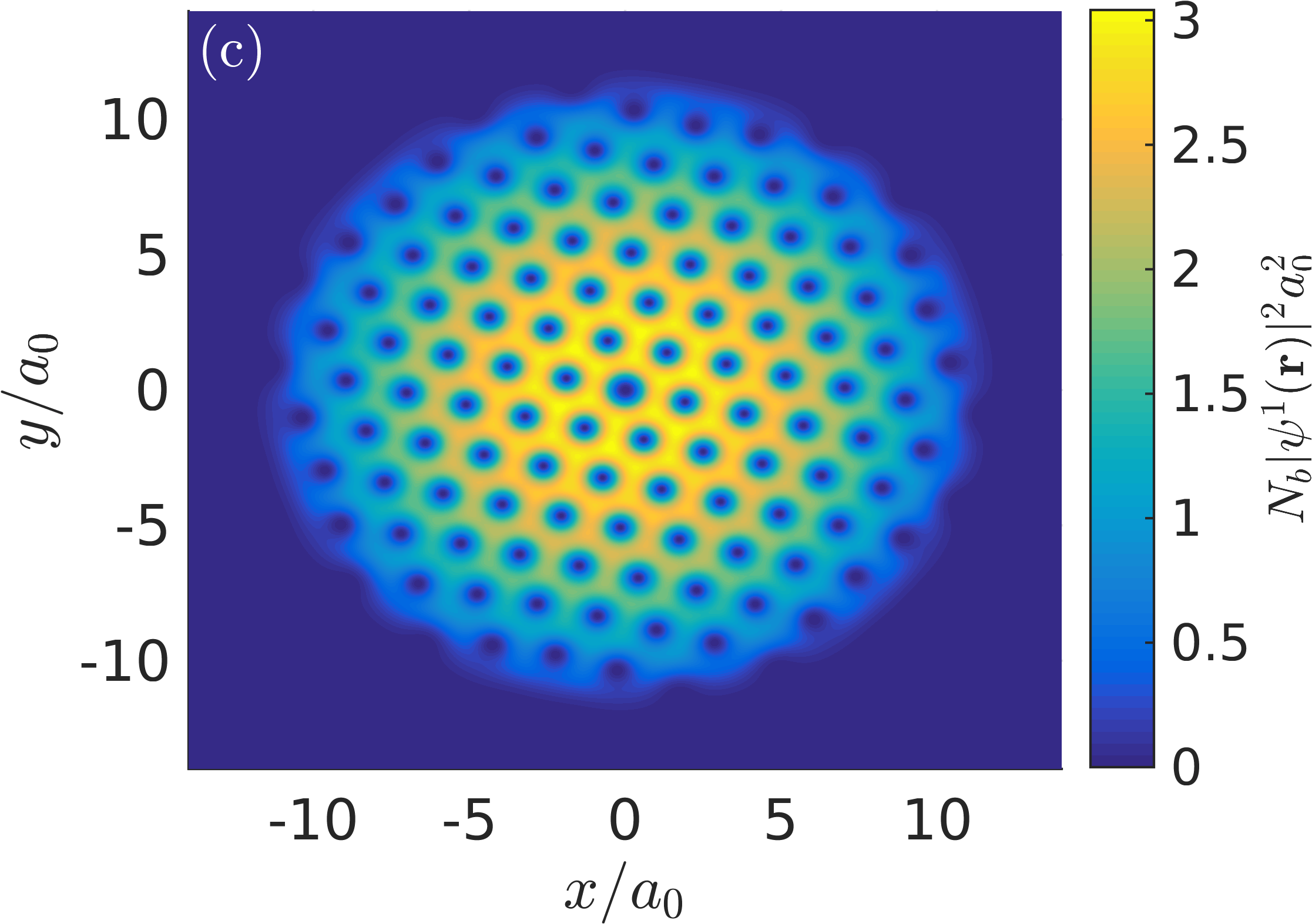}
                \label{fig:becdens1}
        \end{subfigure}
        \hspace{7pt}
        \begin{subfigure}[b]{2.2 in}
                \includegraphics[width=\textwidth]{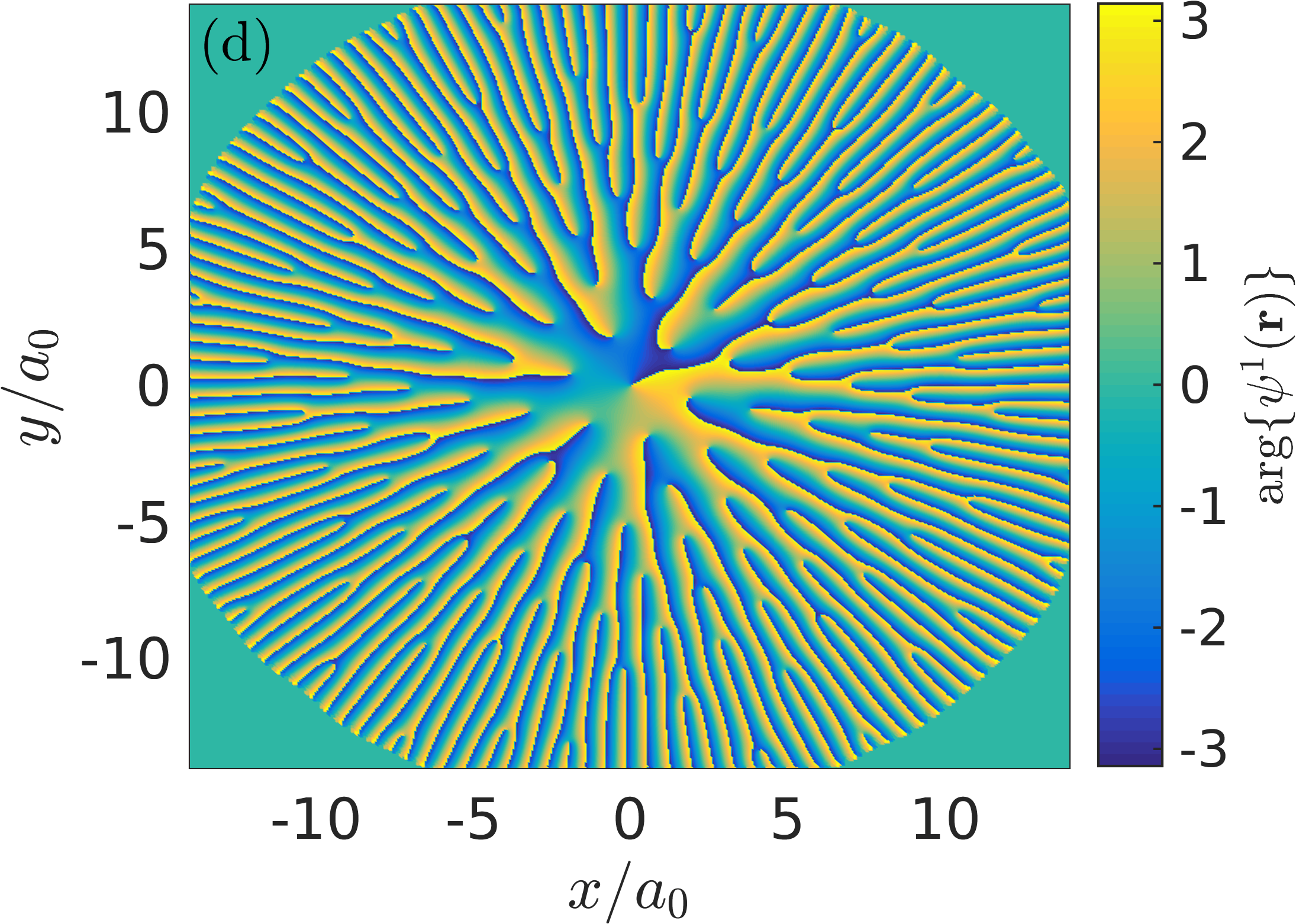}
                \label{fig:becphase1}
        \end{subfigure}
        \hspace{7pt}
        \begin{subfigure}[b]{2.2 in}
                \includegraphics[width=\textwidth]{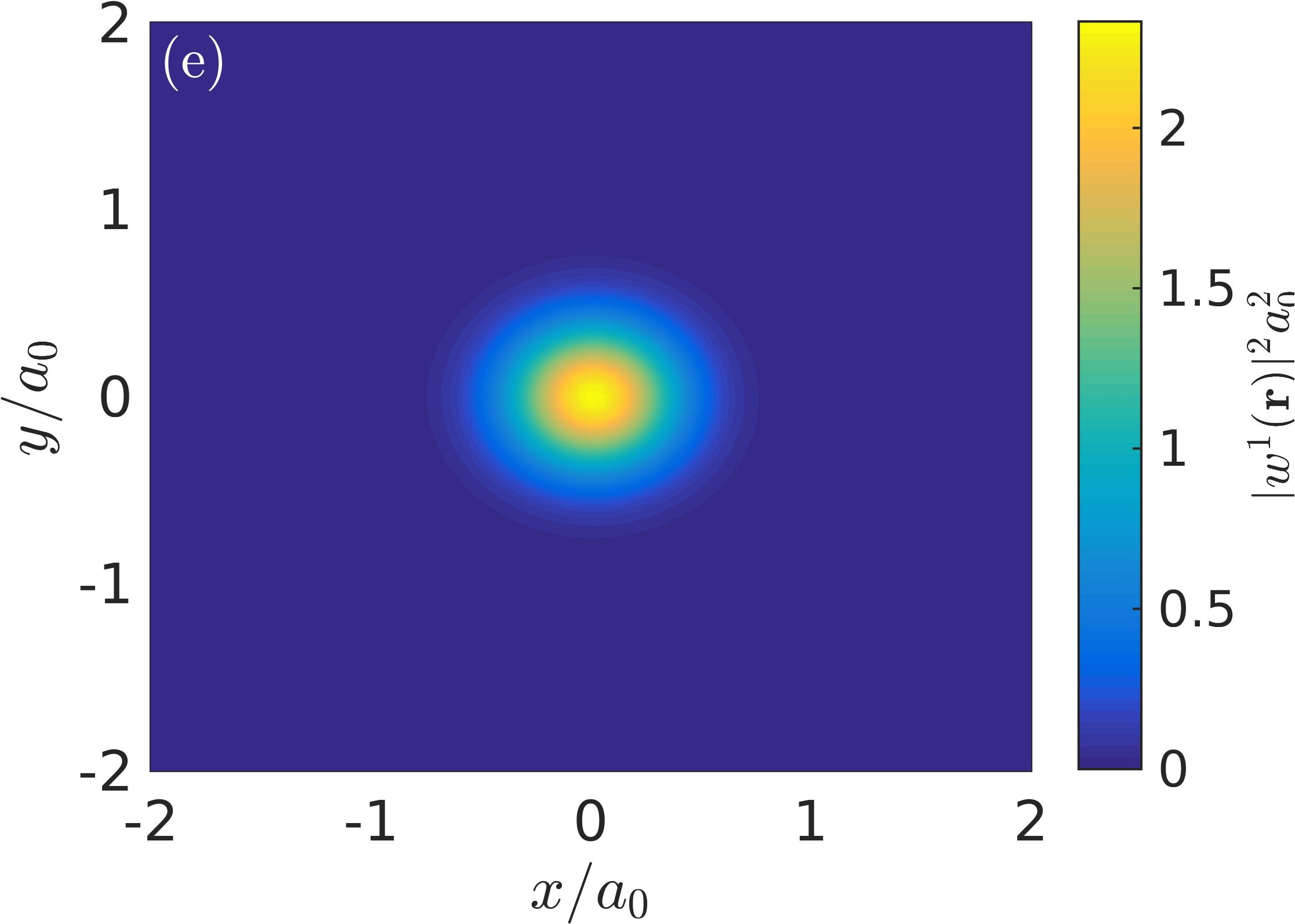}
                \label{fig:impdens1}
        \end{subfigure}
        \caption{\label{fig:fullnumericsspatialplots} {\em Ground state calculations}.  The (a) density $N_b |\psi^{0}(\rr)|^2$ and (b) phase $\arg \{ \psi^{0}(\rr) \}$ of the BEC ground state when there is no impurity present $n = 0$. (c) and (d) show the equivalent quantities, $N_b |\psi^{1}(\rr)|^2$ and $\arg \{\psi^{1}(\rr) \}$, when there is a single impurity $n = 1$ localized at the central vortex. (e) The density $w^1 (\rr)$ of the impurity (its phase is uniform). We use values $N_b = 500$, $\gamma = 1$, $\Omega' = 0.98$, $\alpha = 1$, and $\eta_{ab} = 6$ for the parameters defined following \eqr{eq:lastbeforedefs}.}
\end{figure}


We now solve the above equations using the normalized gradient flow method via the backward Euler Fourier pseudospectral discretization~\cite{Bao2004,Bao2006,Bao2013}. Figures \ref{fig:fullnumericsspatialplots}(a) and (b) show the density and phase of the solution $\psi^0(\rr)$ for $N_b = 500$ bosons with an intermediate interaction strength $\gamma = 1$ and no impurities $n = 0$. A fast rotation $\Omega' = 0.98$ means that the density profile (ignoring vortex cores) is significantly altered from that predicted for a non-rotating BEC. Working within the Thomas-Fermi approximation and making the additional approximation $\nablaa \psi^0(\rr) = \ii m_b \vv (\rr) \psi^0 (\rr) /\hbar$, where $\vv(\rr) = \Omega \hat{\zz} \times \rr$ is the velocity of an equivalent rotating rigid body, we find that the BEC has radius $R$ and bulk density  $n_0 = N_b |\psi^0 (0) |^2$ away from the vortices satisfying~\cite{Fetter2009}
\begin{align}
 \frac{R}{a_0}  = \left ( \frac{4 (N_b - 1) \gamma}{\pi}  \right)^{1/4} \left[ 1 - \left(\frac{\Omega }{ \Omega_b} \right)^2 \right]^{-1/4} \approx 11.3 , \label{eq:TFradius} \\
n_0 a^2_0 =  N_b \left( \frac{1}{\pi (N_b - 1) \gamma} \right)^{1/2} \left[ 1 - \left(\frac{\Omega }{ \Omega_b} \right)^2 \right]^{1/2} \approx 2.51.
\end{align}
These numbers match our numerical results shown in \fir{fig:fullnumericsspatialplots}(a), displaying both a widening and flattening of the condensate by factors $[ 1 - (\Omega / \Omega_b )^2 ]^{-1/4}$ and $[ 1 - (\Omega / \Omega_b )^2 ]^{1/2}$, respectively, due to the rotation. 

The bulk of the system thus starts to resemble the ideal limit that we wish to consider: the BEC density, with the exception of the vortex cores, is essentially homogeneous relative to (varies over a larger length scale than) the other length scales that determine the physics of the system. These two length scales are the distance $a = (2 \pi \hbar / \sqrt{3} m_b \Omega )^{1/2} \approx 1.92 a_0$ between vortices forming a triangular lattice, set by the speed of rotation $\Omega$, and the healing length $\xi = 1 / \sqrt{\gamma n_0} \approx 0.631 a_0$, which controls the size of the vortices and is set by the density in the bulk $n_0$. These values are compatible with the numerically-obtained values for the distance between two central vortices $1.91 a_0$ and the radius of the central vortex $R^0_\mathrm{v} = 0.432 a_0$, where the latter is defined in the caption of \tbr{tb:centralvortex}. We see that, for the parameters chosen, the two relevant length scales $\xi$ and $a$ are roughly on the same order of magnitude, $\xi/a \approx 0.3$, because the rotation $\hbar \Omega$ and interaction energies $n_0 g_b/2$ are similar, $2 \hbar \Omega / n_0 g_b \approx 0.8$. This means that the vortices are tightly packed, with the shape of each vortex core, though largely radially symmetric, significantly affected by the presence of other vortices nearby.

\subsubsection{Impurities in the central vortex}
\label{sec:impinvl}
\begin{table}[tb]
    \begin{tabular}{ p{20pt} | p{10pt} *{4}{p{50pt}} }
    $n$ &  & $R^n_\mathrm{v} /a_0$ & $\sigma^n/a_0$ & $E^n/\hbar \Omega_b$  & $\epsilon^n /\hbar \Omega_b$ \\ \hline \hline
    $0$ &   & $0.4323$ & - & $1383.7$ & - \\
    $1$ &   & $0.5829$ & $0.3660$ & $1392.2$ & $8.4390$ \\
    $2$ &   & $0.6940$ & $0.4017$ & $1400.5$ & $8.4166$ \\
    $3$ &   & $0.7804$ & $0.4306$ & $1409.1$ & $8.4579$\\
    $4$ &   & $0.8464$ & $0.4532$ & $1417.9$ & $8.5650$
\end{tabular}
\caption{ {\em Joint system properties for different numbers $n$ of impurities at the central vortex}. For different $n$ we give the numerically-obtained vortex radius $R^n_\mathrm{v}$, which is the radius at which half-maximum of the BEC density is achieved, the width $\sigma^n$ of the wavefunction of the impurities defined by $( \sigma^{n} )^2 = \int \dd \rr r^2 |w^n(\rr)|^2$, the energy $E^n$ of the ground state of the combined system, and the effective energy $\epsilon^n = (E^n - E^0)/n$ per impurity for $n$ impurities. We choose values $N_b = 500$, $\gamma = 1$, $\Omega' = 0.98$, $\alpha = 1$, $\eta_a = 1.1$, and $\eta_{ab} = 6$ for the parameters defined following \eqr{eq:lastbeforedefs}.}
\label{tb:centralvortex}
\end{table}

We now introduce some impurities into our system. The goal of our calculations is to build up a picture of the behavior of the impurity wavefunction localized at a single vortex core and the effect of impurities on the profile of the BEC near that vortex. Thus we constrain our impurities so that they may only be localized to a single vortex core, choosing the central vortex core. We do this by introducing a box potential $V_a (\rr)$ that makes it energetically unfavorable for the impurities to have a large density at any other vortex. The particular way we ensure localization at the central vortex only is unimportant; we have checked that the detailed shape and width of the box have negligible effect on the results obtained and the corresponding physics.

The results are shown in Figs.\ \ref{fig:fullnumericsspatialplots}(c), (d) and (e) for a single impurity $n = 1$ of the same mass $\alpha = 1$ of the bosons. The strong repulsive impurity-BEC coupling $\eta_{ab} = 6$ means that a localized impurity is obtained with a size on the order of the vortex core $\sigma^1 = 0.366 a_0 \approx 0.580 \xi$ [\fir{fig:fullnumericsspatialplots}(e) and \tbr{tb:centralvortex}]. The phase of the impurity wavefunction is uniform (not shown). The main effect of the impurity on the BEC is to widen the vortex core in which it sits, increasing its radius from $R^0_\mathrm{v} = 0.4323$ to $R^1_\mathrm{v} = 0.5829$ [Figs.\ \ref{fig:fullnumericsspatialplots}(a) and (c), and \tbr{tb:centralvortex}]. In contrast, as can be seen by comparing Figs.\ \ref{fig:fullnumericsspatialplots}(a) and (b) with Figs.\ \ref{fig:fullnumericsspatialplots}(c) and (d), respectively, the impurity has very little effect on the geometry of the lattice. For example, we find that, when the impurity is added, the distance $a = 1.92 a_0$ between the central vortex and its nearest-neighbor is unchanged within the accuracy of the numerical calculations. This is important because we are thus able to consider a vortex lattice of fixed geometry in the presence of impurities.

This general behaviour continues if we introduce a few more impurities. The key properties of the system are presented in \tbr{tb:centralvortex} for $n = 0, 1, \dots, 4$. We see that as the number of impurities is increased the vortex core widens because of the repulsive interaction. The width of the impurity wavefunction also increases, partly due to the widening of the vortex and partly due to the repulsive interactions of the impurities with each other. 

\subsubsection{Energy per impurity}
\label{sec:eperimp}
The energy due to the addition of an impurity localized at a vortex core determines the local parameters of the Hubbard model, as explained in \secr{sec:HubbardNO}. For this reason we have calculated the energy of our system for different numbers of impurities localized at the central vortex core. In \tbr{tb:centralvortex}, we give the ground state energies $E^n = E[w^n,\psi^n]$ [\eqr{eq:energy2}] of the whole impurity-BEC system for numbers $n$ of impurities localized at the central vortex. More insightful are the effective energies per impurity
\begin{align}
\epsilon^n = \frac{E^n - E^0 }{ n } ,
\end{align}
that can be obtained from these total energies $E^n$, also given in \tbr{tb:centralvortex}. Note that this is merely the inverse of the relation $E^n = E^0 + \epsilon^n n$.

There are two main physical processes that affect how the energy per impurity $\epsilon^n$ changes with $n$. The first is the direct repulsive interactions between impurities. All other things equal, this would cause the total energy $E^n$ to increase faster than linearly with increasing $n$ and the energy per impurity $\epsilon^n$ to increase with $n$. This is despite the impurity wavefunction widening and reducing the impurity-impurity interaction energy. The second is the lowering of the energy of each impurity due to the widening of the vortex core caused by itself and the other impurities, which has a similar effect on energy as an effective attractive interaction between impurities. All other things equal, this would then mean the total energy $E^n$ increases slower than linearly with increasing $n$ and the energy per impurity $\epsilon^n$ decreases with $n$. 

The dependence of the energy per impurity $\epsilon^n$ on $n$ is thus the result of these two competing processes, both of which depend separately on the occupation $n$. Both extremes are possible: monotonically increasing or decreasing energy per impurity $\epsilon^n$ with $n$. Indeed, in \secr{sec:HubbardO}, we found that, for weak interactions, only these two extremes are possible [\eqr{eq:epsilonBog}].  Interestingly, for strongly-interacting particles, a non-monotonic energy per impurity $\epsilon^n$ is possible and it is such an example that is shown in \tbr{tb:centralvortex}. Initially, for few impurities $n = 2$, it is the effective attractive interaction that is dominant, but then for more impurities $n > 2$ the direct repulsion is more significant, meaning that $\epsilon^n$ first decreases and then increases with $n$. This can be understood in the following way. The density of the BEC near the vortex can only be lowered so much before it becomes zero. Thus additional impurities will deform the BEC density less and less. This means that while the effect of impurities attracting each other through the BEC deformation can initially be more significant for a small number of impurities, direct repulsive interactions dominate eventually.

\subsection{Single vortex core in the bulk}
\label{sec:singlevortex}
We have established the stability of the lattice geometry in the presence of impurities and some of the properties of an impurity localized in a vortex core in the bulk [\secr{sec:vloverview}] through intensive calculations featuring a full vortex lattice. We now use this knowledge to examine the same physics in the simpler setting of a single vortex core. Our aim is to calculate, in this setting, the shape of the impurity and BEC wavefunctions near the core, and the energy due to the presence of impurities at the core.

\subsubsection{Reduction to a single unit cell}
We begin by making the approximation that near the vortex core, the BEC and impurity wavefunctions take the following symmetric forms [\fir{fig:fullnumericsspatialplots}]
\begin{align}
\psi^n (\rr) =&  \tilde{\psi}^n(r) \ee^{\ii \phi} , \\
w^n(\rr) =& \tilde{w}^n(r) .
\end{align}
We saw in \secr{sec:vloverview} that in the bulk, the external trapping potentials have a negligible effect on the impurities and bosons, thus we do not include such trapping. Instead, we saw that the physics of the BEC is controlled by two length scales: the vortex lattice parameter $a = (2 \pi \hbar / \sqrt{3} m_b \Omega )^{1/2}$ induced by rotation, and the healing length $\xi = 1/\sqrt{\gamma n_0}$ associated with the bulk density $n_0$ of the BEC. We introduce the first length scale by considering a circular domain $\RRR = \{\rr | r \leq \ell \}$ of radius $\ell$ such that its area $\pi \ell^2 = \sqrt{3} a^2 / 2$ (equivalently $\ell = \sqrt{\hbar/m_b \Omega}$) is the same as that of the unit cell of the vortex lattice, and requiring the BEC to reach a maximum at this boundary  
\begin{align}
\od{\tilde{\psi}^n(\ell)}{r} = 0 .
\end{align}
The density, which relates to the healing length, is controlled by fixing the number $N_b$ of bosons in the domain $\RRR$ rather than by fixing the density directly, e.g., at the boundary. 
For clarity, in the remainder of this section, we drop the tildes. 

\subsubsection{The energy functionals and Gross-Pitaevskii equations}
We now insert this simplified ansatz into the energy functionals. The grand canonical energy functional (to be minimized) is written in terms of the energy functional and single-particle Hamiltonians as follows
\begin{align}
E_{\mu_a, \mu_b} [w^n,\psi^n] = &  E[w^n,\psi^n] - 2 \pi \mu_a n  \int_0^\infty \dd rr | w^n (r) |^2 - 2 \pi \mu_b N_b  \int_0^\ell \dd rr | \psi^n (r) |^2 , \\
E[w^n,\psi^n] = &  2 \pi n  \int_0^\infty \dd rr \left( w^n (r) \right)^\ast h_a (r) w^n (r) +  2 \pi \frac{g_a}{2} n (n - 1)  \int_0^\infty \dd rr | w^n (r) |^4  \nonumber \\
& +  2 \pi N_b  \int_0^\ell \dd rr \left( \psi^n (r) \right)^\ast h_b (r) \psi^n (r) + 2 \pi \frac{g_b}{2} N_b^2  \int_0^\ell \dd rr | \psi^n (r) |^4 \\
& + 2 \pi g_{ab} n N_b   \left [ \int_0^\ell \dd rr | w^n (r) |^2 | \psi^n (r) |^2  + \int_\ell^\infty \dd rr | w^n (r) |^2 | \psi^n (\ell) |^2  \right ] , \nonumber
\\
h_a (r) = & -\frac{\hbar^2}{2 m_a} \frac{1}{r} \od{}{r} \left( r \od{}{r} \right), \\
h_b (r) = & \frac{\hbar^2}{2 m_ b r^2} -\frac{\hbar^2}{2 m_b} \frac{1}{r} \od{}{r} \left( r \od{}{r} \right)  - \hbar \Omega .
\end{align}
The corresponding Euler-Lagrange equations, or coupled Gross-Pitaevskii equations
\begin{align}
\left( h_a (r) + g_{ab} N_b | \psi^n (r) |^2 +  g_a (n - 1) | w^n (r) |^2 \right) w^n (r) = & \mu_a w^n (r), \quad \mathrm{for} \: r \leq \ell ,\\
\left( h_a (r) + g_{ab} N_b | \psi^n (\ell) |^2 +  g_a (n - 1) | w^n (r) |^2 \right) w^n (r) = & \mu_a w^n (r),  \quad \mathrm{for} \: r > \ell ,\\
\left( h_b (r) + g_{ab} n | w^n (r) |^2 +  g_{ab} n \frac{\delta(r-\ell)}{2\pi \ell} 2 \pi \int_\ell^\infty \dd r' r' | w^n (r') |^2 +  g_b N_b | \psi^n (r) |^2 \right) \psi^n (r) = & \mu_b \psi^n (r), \quad \mathrm{for} \: r \leq \ell ,
\end{align}
are now much more simply solved, as they feature one-dimensional functions $w^n(r)$ and $\psi^n(r)$ with simpler features. The BEC wavefunction is solved on the reduced domain $0 \leq r \leq \ell$, while the impurity wavefunction is solved in a larger domain. Note that, for the impurity-BEC interaction term, we approximate the BEC density for $r>\ell$ by that at the boundary of its domain $r = \ell$, i.e.\ $|\psi(r)|^2 = |\psi(\ell)|^2$ for $r > \ell$.

\subsubsection{Characteristic quantities}
Before discussing our solution, we again re-express the problem in terms of dimensionless quantities, this time using the domain radius $\ell$ and related rotation energy $\hbar \Omega = \hbar^2 / m_b \ell^2$. So, for example, $r = r' \ell$, $\ell = \ell' \ell$ (i.e.\ $\ell' = 1$), $w^n(r) = w^{n \prime} (r') / \ell$, $E = E' \hbar \Omega$, and $\Omega = \Omega' \Omega$ (i.e.\ $\Omega' = 1$). The result is
\begin{align}
E_{\mu_a', \mu_b'}' [w^{n \prime},\psi^{n \prime}] = &  E'[w^{n \prime},\psi^{n \prime}] - 2 \pi\mu_a' n  \int_0^\infty \dd r' r' | w^{n \prime} (r') |^2 - 2 \pi\mu_b' N_b  \int_0^{1} \dd r' r' | \psi^{n \prime} (r') |^2 , \\
E'[w^{n \prime},\psi^{n \prime}] = & 2 \pi n  \int_0^\infty \dd r' r' \left( w^{n \prime} (r') \right)^\ast h'_a (r') w^{n \prime} (r') + 2 \pi \frac{\eta_a \gamma}{2} n (n - 1)  \int_0^\infty \dd r' r' | w^{n \prime} (r') |^4 \nonumber \\
& + 2 \pi N_b  \int_0^{1} \dd r' r' \left( \psi^{n \prime} (r') \right)^\ast h'_b (r') \psi^{n \prime} (r') +2 \pi  \frac{ \gamma}{2} N_b^2  \int_0^{1} \dd r' r' | \psi^{n \prime} (r') |^4 \\
& + 2 \pi \eta_{ab} \gamma n N_b   \left [ \int_0^{1} \dd r' r' | w^{n \prime} (r') |^2 | \psi^{n \prime} (r') |^2  + \int_{1}^\infty \dd r' r' | w^{n \prime} (r') |^2 | \psi^{n \prime} (1) |^2  \right ] , \nonumber
\\
h'_a (r') = & -\frac{\alpha}{2 } \frac{1}{r'} \od{}{r'} \left( r' \od{}{r'} \right), \\
h'_b (r') = & \frac{1}{2 r^{\prime 2}} -\frac{1}{2} \frac{1}{r'} \od{}{r'} \left( r' \od{}{r'} \right)  - 1 ,
\end{align}
and
\begin{align}
\left( h'_a (r') + \eta_{ab} \gamma N_b | \psi^{n \prime} (r') |^2 +  \eta_a \gamma (n - 1) | w^{n \prime} (r') |^2 \right) w^{n \prime} (r') = & \mu'_a w^{n \prime} (r'), \quad \mathrm{for} \: r' \leq 1 ,\\
\left( h'_a (r')+ \eta_{ab} \gamma N_b | \psi^{n \prime} (1) |^2 +  \eta_a \gamma (n - 1) | w^{n \prime} (r') |^2 \right) w^{n \prime} (r') = & \mu'_a w^{n \prime} (r'), \quad \mathrm{for} \: r' > 1 ,\\
\left( h'_b (r')+ \eta_{ab} \gamma n | w^{n \prime} (r') |^2 +  \eta_{ab} \gamma n \delta(r'-1) \int_1^\infty \dd r' r'| w^{n \prime} (r') |^2 +  \gamma N_b | \psi^{n \prime} (r') |^2 \right) \psi^{n \prime} (r') = & \mu'_b \psi^{n \prime} (r'), \quad \mathrm{for} \: r' \leq 1 .
\end{align}
We see that the physics of the rotation is built into the boundary conditions for the BEC wavefunction. The other quantities characterizing the physics of the system are the relative sizes of interactions $\gamma$, $\eta_a$ and $\eta_{ab}$, defined as before, and the density of the BEC, determined by the number of bosons $N_b$ and controlling the healing length of the system.

\subsubsection{Numerical results}

To allow as much comparison to the earlier calculations as possible, in our examples we set $\gamma = 1$, as before, and $N_b = 10$, which is roughly the number of particles found earlier in a single unit cell. For the impurity, we set $\alpha = 1$ as before. Let us begin by examining the previously-considered values $\eta_a = 1.1$ and $\eta_{ab} = 6$. We solve these simpler coupled Gross-Pitaevskii equations using a Crank-Nicolson finite-difference approach~\cite{Adhikari2000}. 

The BEC and impurity wavefunctions are shown in Fig.~2(a) of the main text. As seen in the earlier calculations featuring multiple vortices, the effect of adding impurities is to widen the vortex core, with the impurity wavefunction spreading out due to this widening and repulsive impurity-impurity interactions. We can also use these calculations to work out the energies $E^n = E[w^n, \psi^n]$ of the unit cell with $n$ impurities localized at it, which in turn define energies per impurity $\epsilon^n = (E^n - E^0)/n$. 

These single-impurity energies are shown in Fig.~2(b) and (c) of the main text for varying $\eta_a$ and $\eta_{ab}$, respectively. As before [\secr{sec:eperimp}], by increasing the importance of repulsive impurity interactions $\eta_{a}$ or decreasing the effective attractive interaction mediated by the deformation of the BEC $\eta_{ab}$ we move from a regime in which $\epsilon^n$ decreases monotonically in $n$ to one in which it increases monotonically. Between these two regimes, there is another in which $\epsilon^n $ first decreases and then increases with $n$. This includes $\eta_a = 1.1$ and $\eta_{ab} = 6$ (for which the effect was witnessed in the full lattice calculations).

\section{Phase diagram in the Gutzwiller ansatz}
\label{sec:Gutzwiller}

\begin{figure}[tb]
        \begin{subfigure}[b]{1.5 in}
                \includegraphics[width=\textwidth]{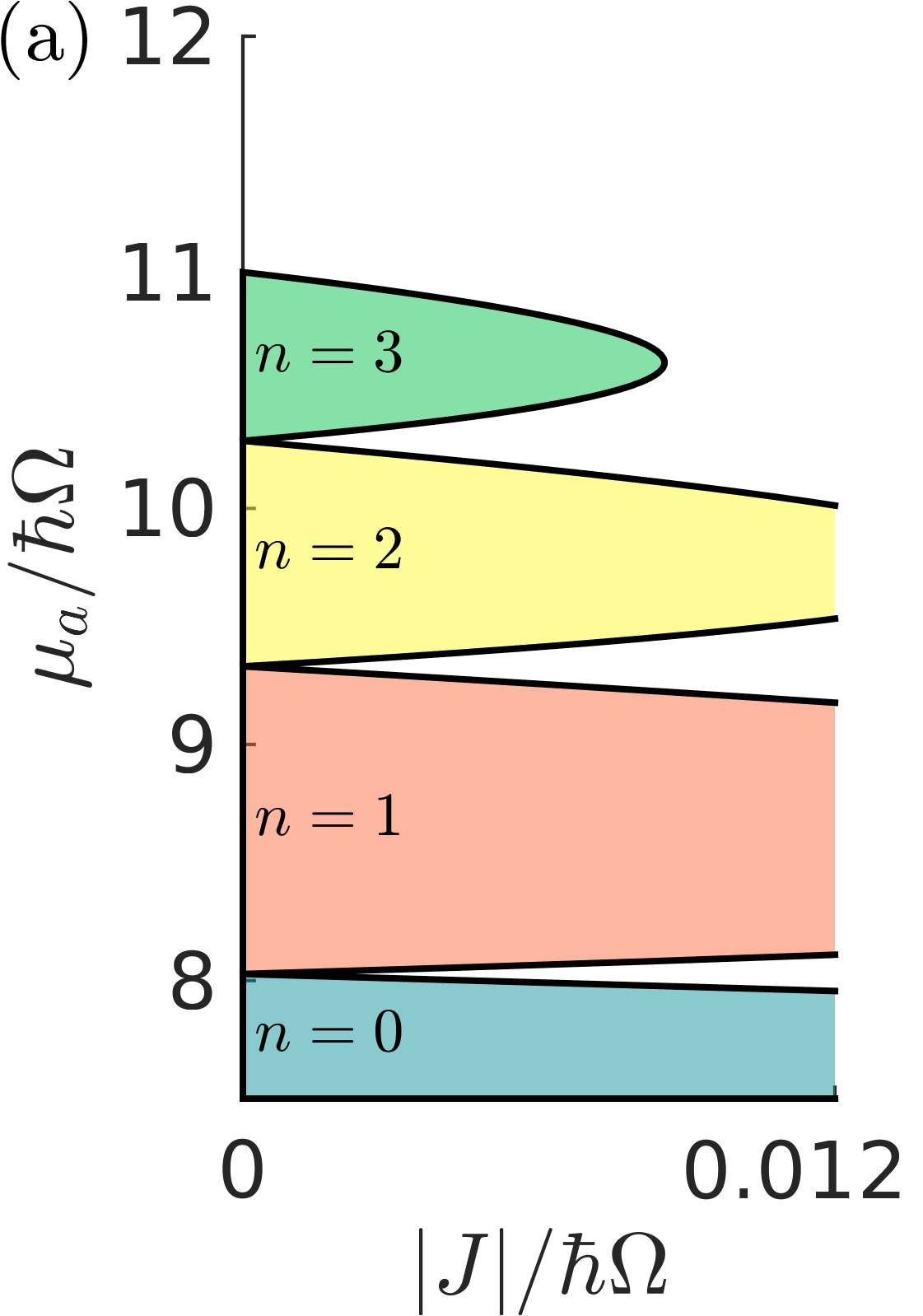}
                \label{fig:wfvsNa}
        \end{subfigure}
        \hspace{7pt}
        \begin{subfigure}[b]{1.5 in}
                \includegraphics[width=\textwidth]{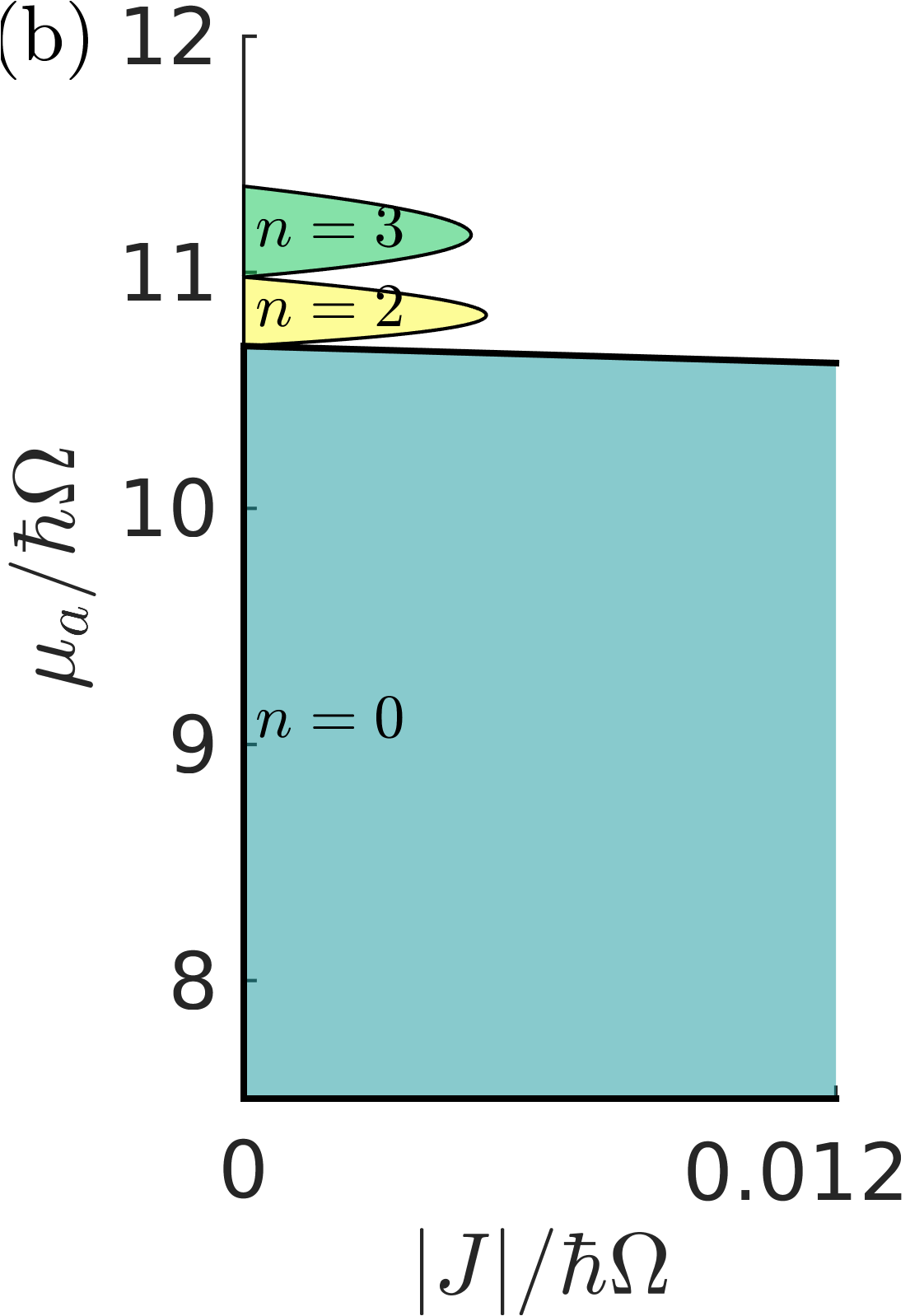}
                \label{fig:energyvsetaa}
        \end{subfigure}
        \caption{\label{fig:phasediagrams} {\em Phase diagram}. The phase diagram for (a) $\eta_{ab} = 3$ and $\eta_a = 1.5$, and (b) $\eta_{ab} = 6$ and $\eta_a = 1.1$. We choose $N_b = 10$ for the number of bosons per unit cell, $\gamma = 1$, and $\alpha = 1$. This is euqivalent to Fig.\ 1(d) in the main text.}
\end{figure}

In this section we calculate the phase diagram of the Hubbard model for the impurities under the assumption of a fixed occupation-independent hopping of magnitude $|J|$. We follow Ref.\ \cite{Umucalhlar2007} and calculate the Mott lobe structure of the phase diagram within the Gutzwiller ansatz for the state.

We work with Hubbard Hamiltonian
\begin{align}
\label{eq:HubbardOagain}
\an{H}_{\mathrm{Hubbard}} = E^0 + \sum_{\RR} \epsilon^{\an{n}_\RR} \an{n}_\RR  + J \sum_{\langle \RR' \RR \rangle} \cre{a}_{\RR'}  \an{a}_\RR ,
\end{align}
and make the unnormalized Gutzwiller ansatz $\ket{G} = \prod_\RR \left( \Delta \ket{n-1} + \ket{n} + \Delta' \ket{n+1} \right)$ for the state of the impurities near the boundary of the $n$-th Mott lobe. This simple ansatz is appropriate when the mass of an impurity equals that of a boson, $\alpha = 1$, since there will be one effective unit of magnetic flux per vortex site and thus unit cell. A more complicated calculation may be performed for a different mass ratio $\alpha \neq 1$, but the resulting phase diagram will be qualitatively similar \cite{Umucalhlar2007}.

Minimizing 
\begin{align}
\label{eq:GutzwillerMin}
\frac{\brackets{G}{ ( \an{H}_{\mathrm{Hubbard}} -\mu_a \sum_\RR \an{n}_\RR )}{G}  }{ \brakets{G}{G} } ,
\end{align}
gives the approximate ground state in the grand canonical ensemble, with impurity chemical potential $\mu_a$. The ground state undergoes a transition from $\Delta = \Delta' = 0$ to $\Delta, \Delta' \neq 0$ as $|J|$ increases beyond
\begin{align}
\label{eq:Jc}
|J|_c = \frac{1}{z} \frac{ ( \epsilon^{n+1} (n+1) - \epsilon^n n - \mu_a  )   (  \epsilon^{n-1} (n-1) - \epsilon^n n + \mu_a  )  }{n ( \epsilon^{n+1} (n+1) - \epsilon^n n - \mu_a  ) + (n+1) (  \epsilon^{n-1} (n-1) - \epsilon^n n + \mu_a  ) } ,
\end{align}
where $z=6$ is the number of nearest neighbours to each site. The transition points for different $|J|$ and $\mu_a$ are plotted in \fir{fig:phasediagrams} as well as in Fig.\ 1(d) of the main text. 

The minimum and maximum chemical potential $\mu_a$ in the $n$-th Mott lobe are given by
\begin{align}
\mu_a = \epsilon^{n} n - \epsilon^{n-1} (n-1) , \\
\mu_a = \epsilon^{n+1} (n+1) - \epsilon^n , 
\end{align}
respectively, which are just the energy differences between different occupations of the site. A lobe is missing if this maximum falls below the minimum i.e.\ if
\begin{align}
\epsilon^{n+1} (n+1) + \epsilon^{n-1} (n-1)- 2 \epsilon^n  < 0 ,
\end{align}
which corresponds to a negative curvature in the total energy $E^n = E^0 + \epsilon^n n$ as a function of $n$ impurities at a site. For our example in \fir{fig:phasediagrams}(b) this has occurred for the $n=1$ Mott lobe.

\section{Tunneling between two wells}
\label{sec:tunnelling}

\begin{figure}[tb]
        \begin{subfigure}[b]{2.2 in}
                \includegraphics[width=\textwidth]{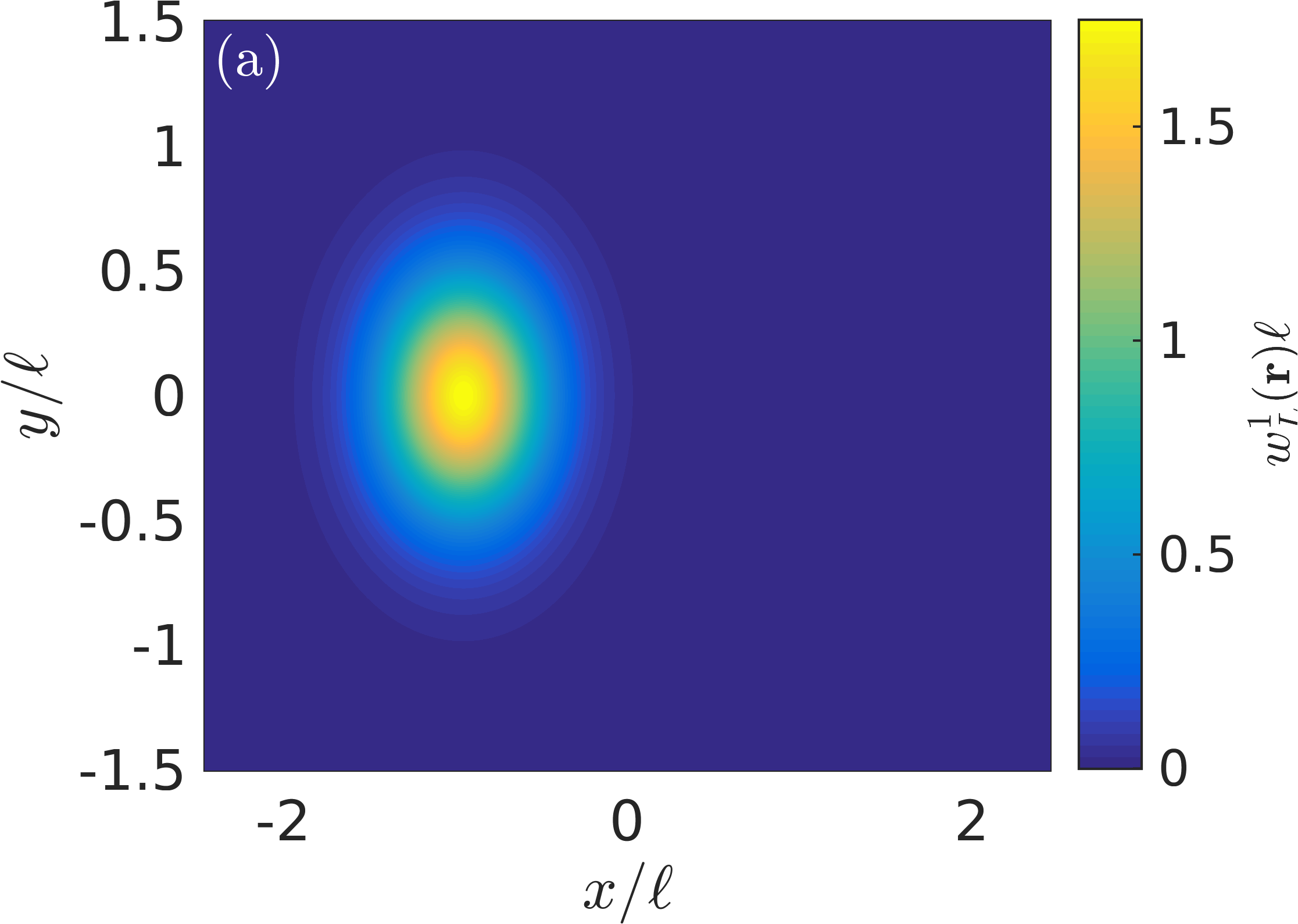}
                \label{fig:wfvsNa}
        \end{subfigure}
        \hspace{7pt}
        \begin{subfigure}[b]{2.2 in}
                \includegraphics[width=\textwidth]{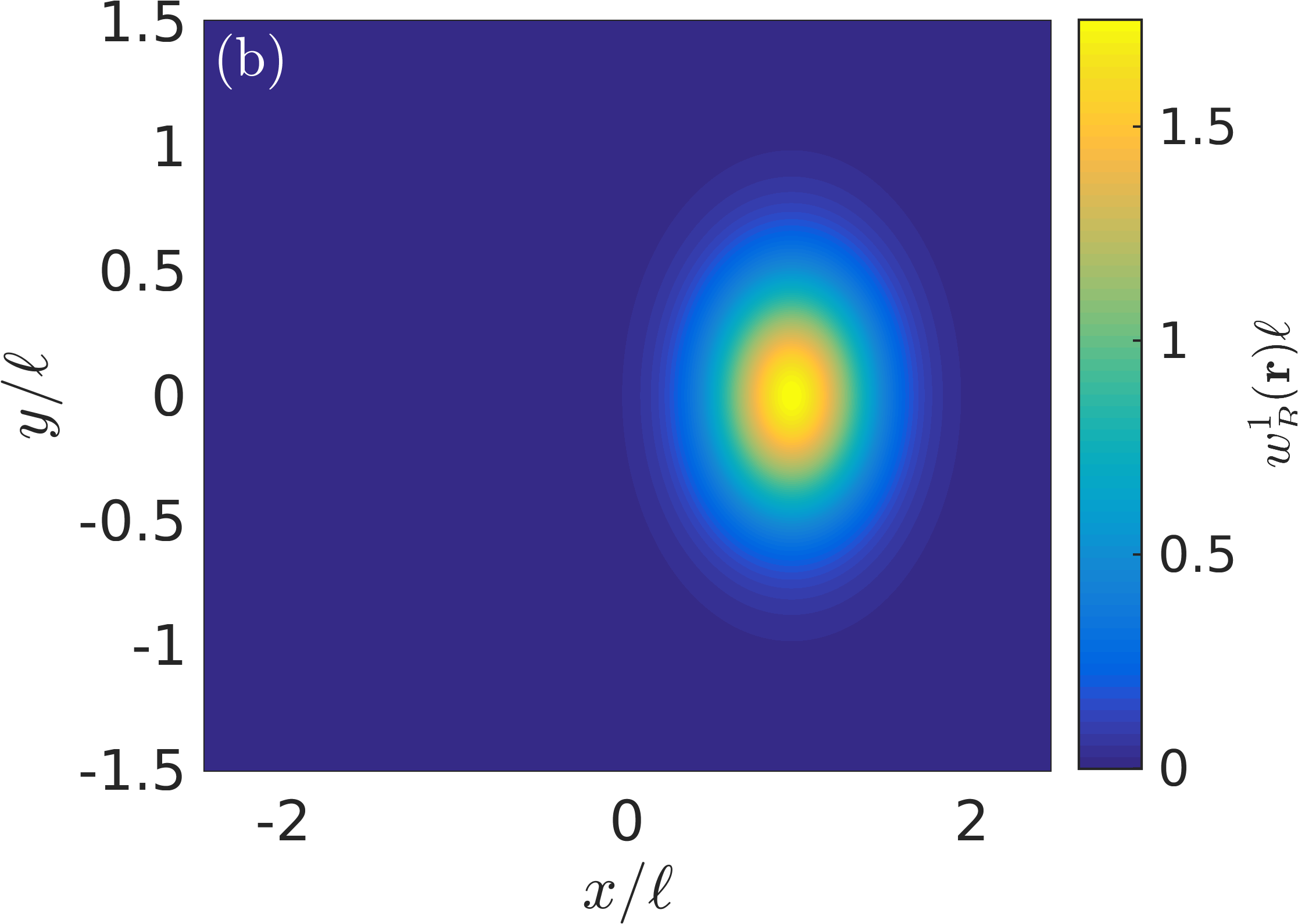}
                \label{fig:energyvsetaa}
        \end{subfigure}
        \hspace{7pt}
        \begin{subfigure}[b]{2.2 in}
                \includegraphics[width=\textwidth]{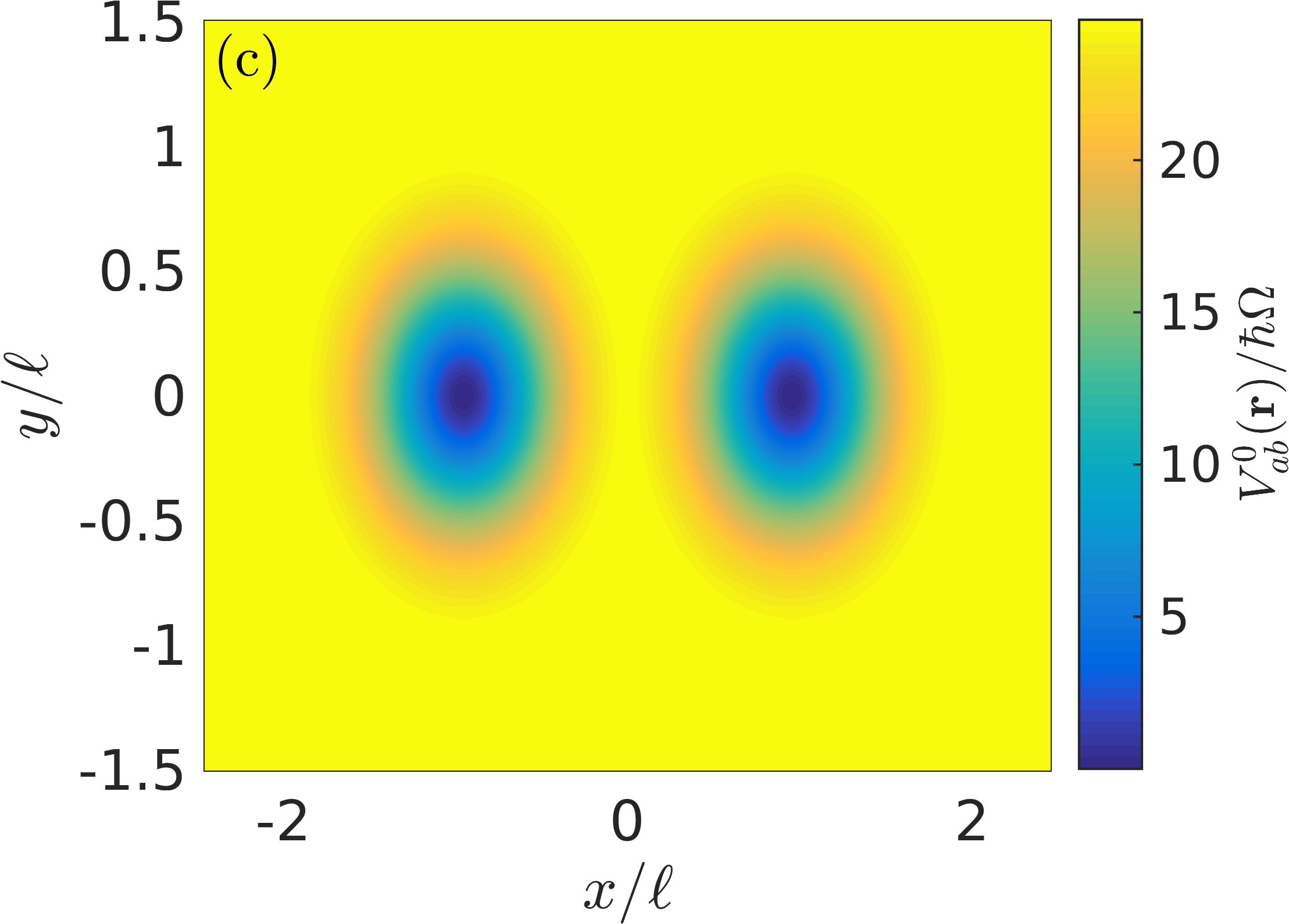}
                \label{fig:energyvsetab}
        \end{subfigure}
        \caption{\label{fig:tunnelingplots} {\em Tunneling between two wells}. (a) and (b) The orthogonalized single-impurity well function at the left $w^1_L (\rr)$ and right $w^1_R (\rr)$ well. (c) The potential $V^0_{ab} (\rr) = g_{ab} N_b |\psi^0 (\rr) |^2$ due to the unperturbed BEC. We choose $N_b = 10$ for the number of bosons per unit cell, $\gamma = 1$, $\alpha = 1$ and $\eta_{ab} = 6$.}
\end{figure}


In this section, we estimate the size of the hopping parameter $|J|$ to assess the region of the phase diagram that our system occupies. In \secr{sec:HubbardNO} we outline how these parameters might be estimated with more accuracy but, since we only require an estimate of the magnitude, we follow a simple heuristic approach. We first estimate the unrenormalized hopping, associated with an impurity hopping in the potential $V^0_{ab} (\rr) = g_{ab} N_b |\psi^0 (\rr) |^2$ created by the unperturbed BEC. Then we argue that the true hopping, once renormalized, will be smaller than this.

We begin by taking the solutions $w^1 (r)$ and $\psi^0(r)$, the single-impurity well function and unperturbed vortex core, from \secr{sec:singlevortex}. We extrapolate these functions to build objects on a two-dimensional domain containing two vortex cores/sites separated by the lattice parameter $a$. Specifically, we use $w^1 (r)$ to create two impurity well functions $w^1_L (\rr)$ and $w^1_R (\rr)$ at a left and right site, denoted by $L$ and $R$, respectively. The two well functions are orthogonalized i.e.\ mixed slightly such that their overlap is zero. The resulting functions are shown in Figs.\ \ref{fig:tunnelingplots}(a) and (b). The BEC unperturbed vortex core wavefunction $\psi^0(r)$ is used to construct an unperturbed BEC density $|\psi^0 (\rr) |^2$ over the two sites, giving an approximate potential $V^0_{ab} (\rr)$ that is felt by the impurities, shown in \fir{fig:tunnelingplots}(c).

Using these, we estimate the unrenormalized impurity hopping, neglecting rotations,
\begin{align}
\label{eq:jbare}
j^{1 1}_{L R} = \int \dd \rr \left ( w^1_{L} (\rr) \right)^\ast \left[ -\frac{\hbar^2 \nablaa^2}{2 m_a} + V^0_{ab}(\rr) \right ]  w^1_{R} (\rr)  \approx -0.0074 \hbar \Omega ,
\end{align}
for our usual parameters: $N_b = 10$ for the number of bosons per unit cell, $\gamma = 1$, $\alpha = 1$ and $\eta_{ab} = 6$. 

Rotations introduce a Peierls phase to the hopping, as in \eqr{eq:peierlstrans}. The hopping is also reduced by a renormalization factor of the non-unity overlap between the BEC state before and after the transition, as with \eqr{eq:renormalizing} for weak interactions. For strong interactions and multiple impurities per site, this effect is more prominent and there are additional renormalization factors due to similar changes in the wavefunctions of the non-transitioning impurities. Further, by using the unperturbed potential $V^0_{ab} (\rr)$ in our calculation, rather than the deformed potentials, we overestimate the hopping.

In summary, we expect that for the parameter regimes we consider that the hopping energy $|J|$ will be an order of magnitude smaller than the other energy scales in the system e.g.\ the differences between the per-impurity energies $\epsilon^n$. Comparing with \fir{fig:phasediagrams}(b), we find that the system is close to the strongly interacting limit and expect that it will realize a Mott insulator, with a filling determined by the energy per impurity $\epsilon^n$.

\bibliography{paper}

\end{document}